\documentclass[twocolumn]{aastex631}
\usepackage{CJK}

\graphicspath{{./}{figures_COC/}{figures_SGPS/}}
\begin{document}
\begin{CJK*}{UTF8}{gbsn}
\title{A New Tidal Stream Discovered in Gaia DR3}
\correspondingauthor{Hao Tian}
\email{tianhao@nao.cas.cn}
\author[0000-0003-3347-7596]{Hao Tian }
\affil{Key Laboratory of Space Astronomy and Technology, 
National Astronomical Observatories, 
Chinese Academy of Sciences, 
Beijing 100101, P.R. China}
\affil{Institute for Frontiers in Astronomy and Astrophysics, Beijing Normal University, 
Beijing, 102206, People's Republic of China}
\author[0000-0002-1802-6917]{Chao Liu }
\affil{Key Laboratory of Space Astronomy and Technology, 
National Astronomical Observatories, 
Chinese Academy of Sciences, 
Beijing 100101, P.R. China}
\affil{Institute for Frontiers in Astronomy and Astrophysics, Beijing Normal University, Beijing, 102206, People's Republic of China}

\author[0000-0001-8799-480X]{Changqing Luo }
\affil{Key Laboratory of Space Astronomy and Technology, 
National Astronomical Observatories, 
Chinese Academy of Sciences, 
Beijing 100101, P.R. China}

\author[0000-0002-0642-5689]{Xiang-Xiang Xue }
\affil{Key Laboratory of Optical Astronomy, 
National Astronomical Observatories, 
Chinese Academy of Sciences, 
Beijing 100101, P.R. China}
\affil{Institute for Frontiers in Astronomy and Astrophysics, Beijing Normal University, 
Beijing, 102206, People's Republic of China}

\author[0000-0002-3180-2327]{Yujiao Yang }
\affil{Key Laboratory of Space Astronomy and Technology, 
National Astronomical Observatories, 
Chinese Academy of Sciences, 
Beijing 100101, P.R. China}

\begin{abstract}

Thanks to the precise astrometric measurements of proper motions by the Gaia mission, a new tidal stellar stream has been discovered in the northern hemisphere. The distribution of star count  shows that the stream is  approximately $80^\circ$   long and  $1^\circ.70$ wide. Observations of
 $21$ member stars, including 14 RR Lyrae stars, indicate that the stream has an eccentric and retrograde orbit with $e=0.58$. The low metallicity, high total energy, and large angular momentum  suggest that it is associated with the merging event Sequoia. 
This discovery suggests the possibility of finding more substructures with high eccentricity orbits, even in the inner halo. 
\end{abstract}

\section{Introduction} \label{sec:intro}
As a relic of merging events, stellar streams keep the memory of their progenitors' chemical and dynamical information, even after a few giga years \citep{Helmi2020ARA&A..58..205H}.
Many cold streams have been discovered in photometric surveys using the matched-filter 
method \citep{Rockosi2002AJ....124..349R, Grillmair2006ApJ...643L..17G, Grillmair2009ApJ...693.1118G, 
Grillmair2006ApJ...645L..37G, Grillmair2006ApJ...639L..17G,
Grillmair2014ApJ...790L..10G,Grillmair2013ApJ...769L..23G,Bernard2016MNRAS.463.1759B}.
Most of those cold streams are formed from stripped globular clusters that have smaller velocity dispersion.
After the second data release of the Gaia mission, many such cold streams were discovered using proper motions \citep{Malhan2018MNRAS.477.4063M,Grillmair2022ApJ...929...89G}. 
Besides those cold streams,  clearer views of a few broad or diffuse substructures are also shown, such as the Sagittarius Stream and the Virgo Overdensity. These substructures are
believed to be formed through the stripped dwarf galaxies. 
All these substructures show us a strongly disturbed Milky Way halo. Moreover, the accurate astrometric data from Gaia Mission also reveals a lot of phase mixed debris in the phase space 
\citep{Koppelman2018ApJ...860L..11K,Koppelman2019A&A...631L...9K}, especially the largest 
merging event in the Milky Way, 
 Gaia-Enceladus-Sausage  \citep{Helmi2018Natur.563...85H,Belokurov2018MNRAS.478..611B},
whose debris dominates the inner halo. 

With different methods, there are more than a hundred streams, most of which are cold streams, discovered by now \citep{Mateu2018MNRAS.474.4112M}. 
What should be noticed is that many of those streams have not been yet confirmed 
with spectroscopic observations.
The new-discovered broad or  diffuse streams, on the other hand, are not increased, since they are associated 
with more intensive but less-frequently occurred mergers and often with lower surface number density. 
This means that they are much more difficult to  discover
from the field stars than the cold streams. 

In this work, we use Gaia DR3 data to search substructures in the halo with clean removal of field star contamination. We introduce the data selection in Section 2. The density distribution of the discovered stream is shown in Section 3.
The properties of the newly discovered stream are studied in Section 4. Finally, the conclusions are drawn in Section 5.

\section{Sample Selection\label{sec:data}}
With Gaia DR3 data 
\citep{GDR32022arXiv220800211G}, including accurate parallaxes and proper motions, 
we can search substructures in the stellar halo. 
We select the distant stars with accurate proper motions satisfying the following criteria,
 \begin{enumerate}
 \item[1.]  $\omega<0.1$ and $\omega/\sigma_\omega<1$,
 \item[2.]  $\sigma_{\mu_\alpha^*}<0.2$ mas yr$^{-1}$ and  $\sigma_{\mu_\delta}<0.2$ mas yr$^{-1}$,
 \item[3.]  RUWE$<1.2$.
 \end{enumerate}
The constraints on the parallax $\omega$ are used to remove 
the majority of nearby stars ($d<10$ kpc), most of which are disk populations.
This  helps us significantly reduce the  foreground stars and clarify the signal of the substructures, especially the faint ones.
The stars with larger uncertainties in the proper motions are also removed. 
The constraint on  RUWE
 is used to remove those possible binary stars. 
All those selections leave the samples mainly brighter than $G=19$. The  contamination
of galaxies in the faint magnitudes will not affect our results because the proper motions
of the galaxies are around 0 with a dispersion of around $0.2$ mas yr$^{-1}$.

\section{Slicing the proper motion and density distribution}
\begin{figure*}
    \centering
    {\bf View from the North Galactic Pole}\\ \vspace{0.2cm}
    \includegraphics[trim={7cm 0 6.7cm 0},clip,width=0.33\textwidth]{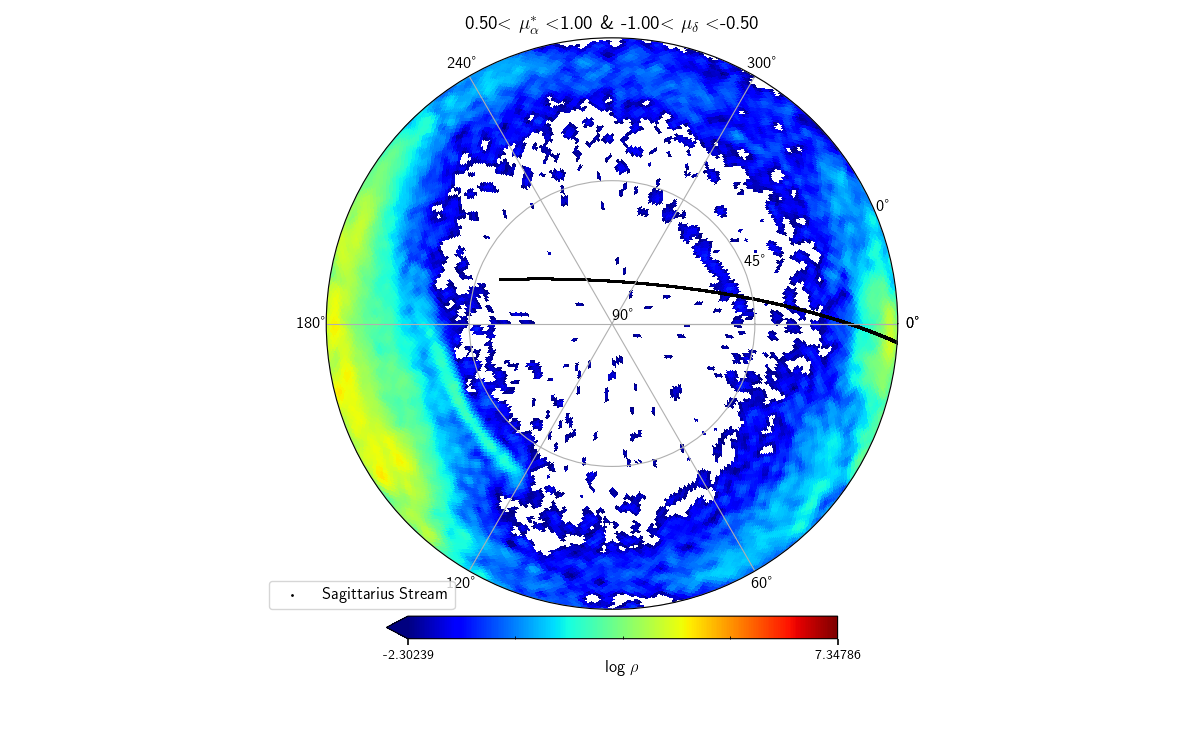} \hspace{-0.5cm}
    \includegraphics[trim={7cm 0 6.7cm 0},clip,width=0.33\textwidth]{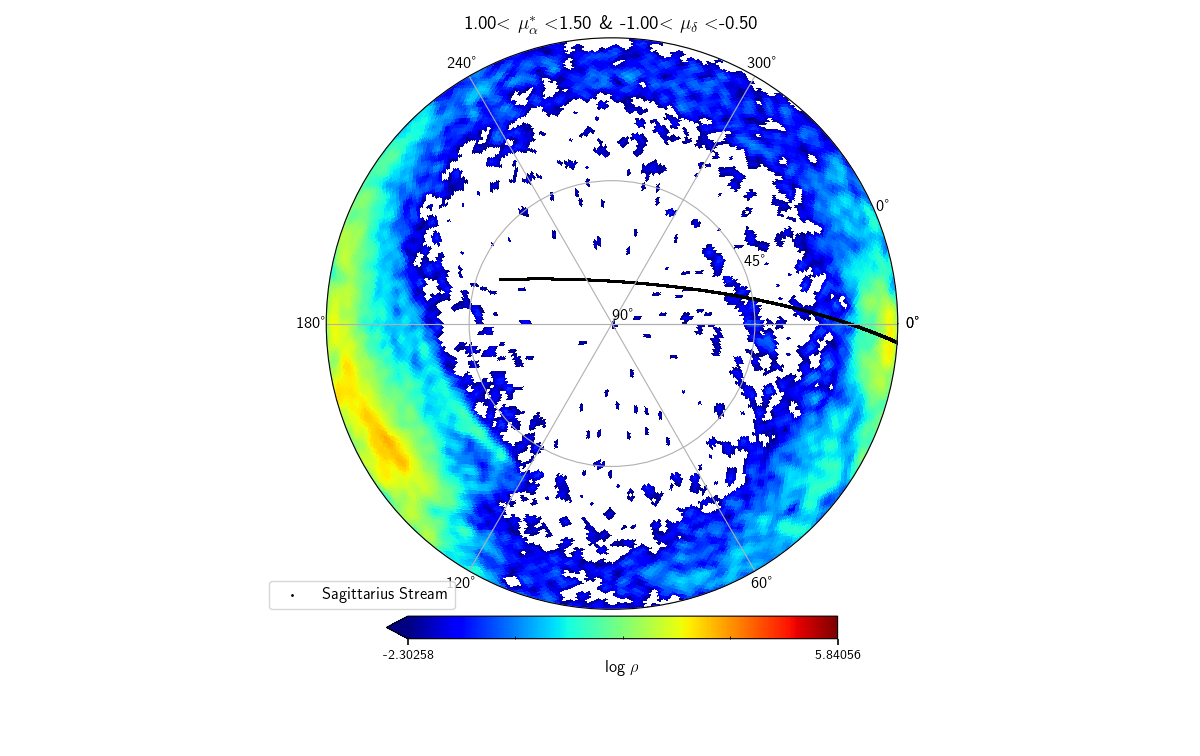} \hspace{-0.5cm}
    \includegraphics[trim={7cm 0 6.7cm 0},clip,width=0.33\textwidth]{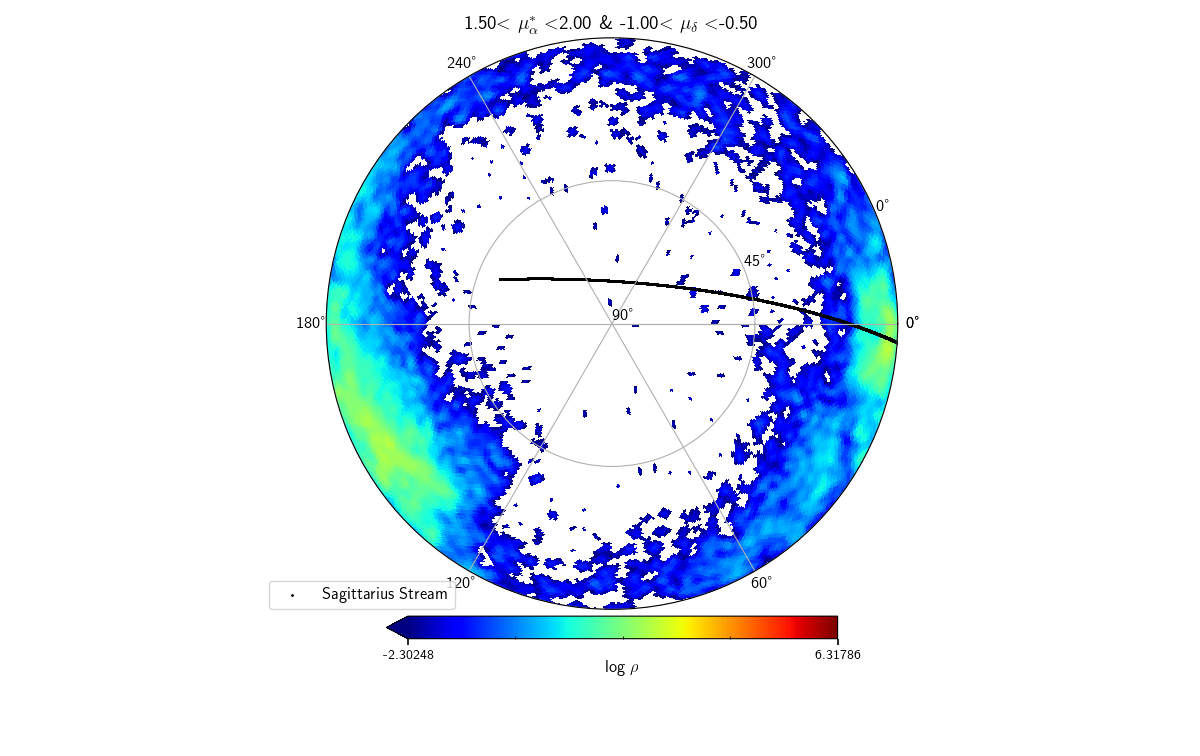} \\ \vspace{-0.2cm}
    

     \caption{
     The density distributions of three subsamples from Gaia DR3 
     with different proper motion ranges are shown with view from
     the north galactic pole. The newly discovered stream is located on the top right in each panel with $270^\circ<l<30^\circ$ and $30^\circ<b<60^\circ$.
     The orbit of the Sagittarius
     Stream \citep{Mateu2018MNRAS.474.4112M,Antoja2020A&A...635L...3A} 
     is represented with 
     solid black lines.
     }
    \label{fig:Density}
\end{figure*}
In principle, the member stars of a unique substructure
in a small space volume will have similar positions and velocities.
Then their proper motions should be also similar.
In other words, the spatial overdensity of the stars with 
similar proper motions should be of high probability to be a substructure from the same progenitor.
We therefore check the spatial density distributions with different 
proper motion ranges in the selected samples of Gaia DR3. The step of $0.5$ mas yr$^{-1}$ is adopted 
to split out the data in both of the proper motions 
$\mu_\alpha^*$ and $\mu_\delta$.  The sky is divided into equal-area pieces by the package
 \texttt{Healpy} with \textit{nside}$=64$ \citep{healpy_2005ApJ...622..759G,healpy_Zonca2019}. 
 The smoothing process is applied using a Gaussian kernel with $\sigma\sim55'$. 
 With these procedures, many known substructures are
 revealed in the density distributions. All those substructures, in form of overdensities 
 in the spatial density distribution, are cross-matched with the latest list of 
 known nearby streams collected by \cite{Mateu2018MNRAS.474.4112M}.
 
As seen in Figure~\ref{fig:Density}, the Anticenter Stream is the most significant 
substructure in the spatial density distribution with galactic
longitude between $120^\circ$ and $200^\circ$ \citep{Laporte2020MNRAS.492L..61L}. 
This proves that our method works well to reveal the substructures. Meanwhile, the orbit of the Sagittarius Stream is indicated by the solid black line \citep{Mateu2018MNRAS.474.4112M, Antoja2020A&A...635L...3A}.

\subsection{Discovery of the new stream}
A new $80^\circ$ long stream with a high signal-to-noise ratio is discovered with  $-1.0<\mu_\delta<-0.5$ mas yr$^{-1}$ and $0.5<\mu_\alpha^*<2.0$ mas yr$^{-1}$ , which is shown  in Figure~\ref{fig:Density} viewed from the Galactic North Pole. 
The new stream is significantly located at a high galactic latitude around $30^\circ$ to $60^\circ$
with galactic longitude from $270^\circ$ to $30^\circ$. 
Different parts of the stream are revealed in different proper motion ranges
because of the projection effect of the Solar motion and the velocity gradient of the stream itself. 
Although the surface density is very low, the stream is still significant compared 
to its neighborhood.
On the right panel in Figure~\ref{fig:Density} with $1.5<\mu_\alpha^*<2.0$ mas yr$^{-1}$, 
 we can still find the extension of the stream following the elongation
 at a lower galactic latitude ($b<45^{\circ}$ and $l\sim30^\circ$). 
 

 \begin{figure*}
    \centering
    \includegraphics[trim={.2cm 0 1.cm 0}, clip, width=0.99\textwidth]{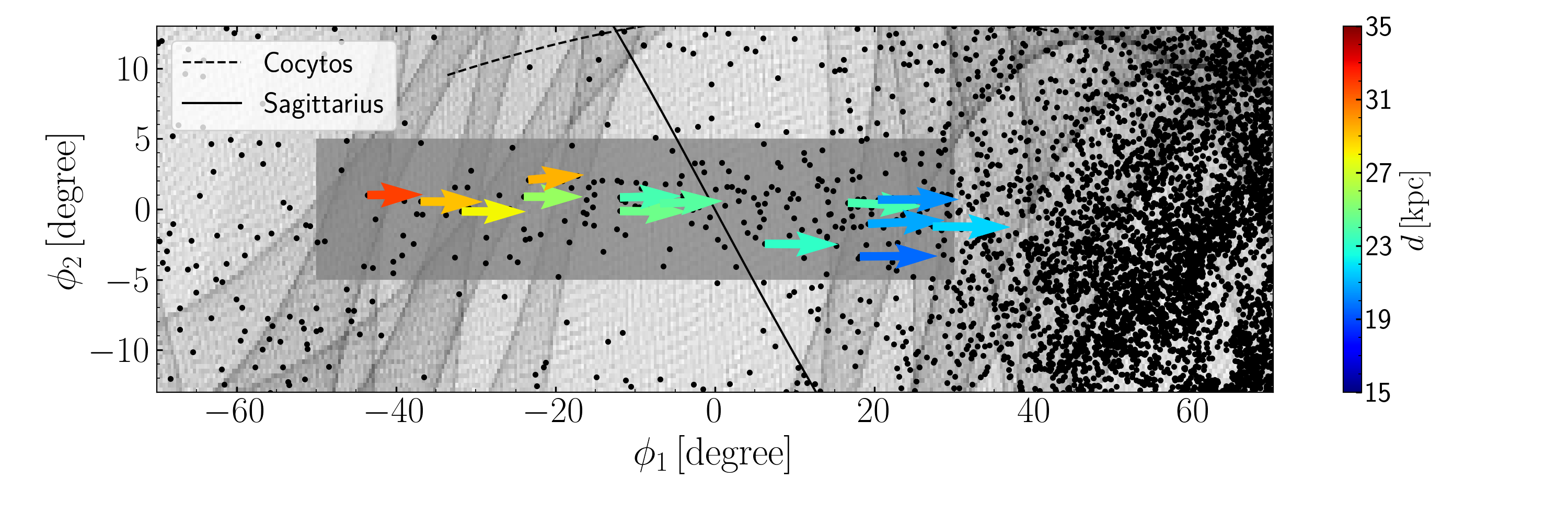} \\
     \caption{ The stars  with $-1.0<\mu_\delta<-0.5$ mas yr$^{-1}$ and $0.5<\mu_\alpha^*<2.0$ mas yr$^{-1}$ are represented by the gray dots in the target region.
     The region of the new stream is marked with gray shadow.
     The orbits of the nearby Cocytos Stream and 
     the Sagittarius Stream are represented by the dashed and dotted lines. The arrows represent
     the tangential velocities of the RR Lyrae stars relative to the position of the Sun, but with 
     the Solar movement corrected. The background represents the scanning law of the Gaia DR3.}
    \label{fig:radec}
\end{figure*}
To avoid the projection effect in the equatorial frame, we convert the coordinate to a new frame,
where the longitude $\phi_1$ is generally along the stream so that the stream is located at the latitude $\phi_2\sim0^\circ$. 
The conversion is done with the function {\it transform\_to} from the package \texttt{gala}\footnote{https://gala.adrian.pw/en/latest/coordinates/greatcircle.html} \citep{gala}.
Two points on the stream are used to define the great circle during
the conversion,  $(\alpha,\delta)=(192.0^\circ,\,-11.5^\circ)$ and $(249.0^\circ,\,4.2^\circ)$. The zero point
is defined as the spherical midpoint of those two points by \texttt{gala} with default setting.
Figure~\ref{fig:radec} shows the field of the streams in the new coordinates, including the orbits of
the Sagittarius Stream \citep{Antoja2020A&A...635L...3A} and 
the Cocytos Stream \citep{Grillmair2009ApJ...693.1118G}, which are represented by 
the solid and dashed lines, respectively. 
 The scanning law of Gaia DR3 is represented in the background, which is the number distribution
 sampled with  an interval of 10 seconds during its 
 observations\footnote{https://gaia.aip.de/metadata/gaiadr3/commanded\_scan\_law/}. That  proves that
the stream is not an artefact relative to the Gaia measurements.

 \begin{figure}
    \centering
    \includegraphics[ width=0.49\textwidth]{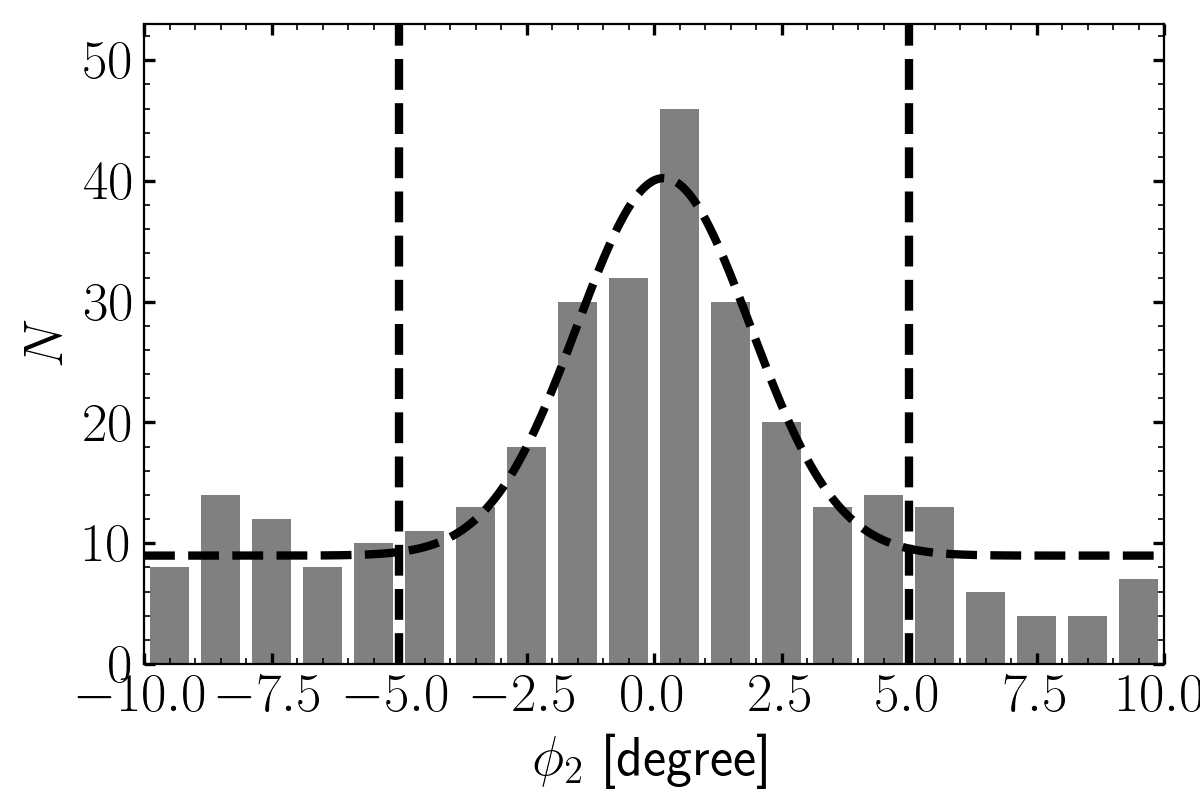}\\
     \caption{
     The distribution of the latitude $\phi_2$ for the stream is shown
     with $-50^\circ<\phi_1<30^\circ$. 
     The thick dashed lines represent the constraints on the latitude 
     for the member selection. The thin dashed line represents a Gaussian 
     fitting results with dispersion of $\sigma_{\phi_2}=1.70^\circ$,
       amplitude of $A=133.48$ and a background of $H=8.98\pm1.1$.
     }
    \label{fig:hist_phi2}
\end{figure}
 Focusing on the new stream, we first select the stars with longitude $\phi_1$ between $-50^\circ$ and $30^\circ$.
 The latitude distribution  $\phi_2$ of the stars is shown in Figure~\ref{fig:hist_phi2}.
 A significant peak is around $\phi_2\sim0^\circ$. To figure out the significance,
 we use the Gaussian function to fit the distribution of $\phi_2$, i.e. $y=A*\mathrm{Gaussian}(\bar{\phi_2},\sigma_{\phi_2})+H$.
 The amplitude of the signal is constrained as $A=133.48$ stars,  the mean value and dispersion
 of the Gaussian distribution are $(\bar{\phi_2},\sigma_{\phi_2})=(0.19^\circ\pm0.16^\circ, 1.70^\circ\pm0.18^\circ)$, and the background
 is $H=8.98\pm1.1$. The fluctuation of the background $\sigma_H$ is calculated by
 the dispersion of the bins
 outside $3\sigma_{\phi_2}$, i.e. $|\phi_2|>5^\circ$. 
 Then the signal-to-noise ratio of the structure is estimated by
 $SNR=A/\sigma_H=39.5$. 
 We take all the $227$ stars 
 within $3\sigma_{\phi_2}$ as the member stars of the stream, the stream has a 
 length of $\sim80^\circ$ and a width of $1.70^\circ$ ($1\sigma_{\phi_2}$).
We also estimate that the contamination of field stars within 10 bins from $\phi_2=-5^\circ$ to $5^\circ$
is about $10*H=90$ in total.

\begin{figure*}
    \centering
  {\bf CMD of member samples} \\
    \includegraphics[height=0.21\textwidth]{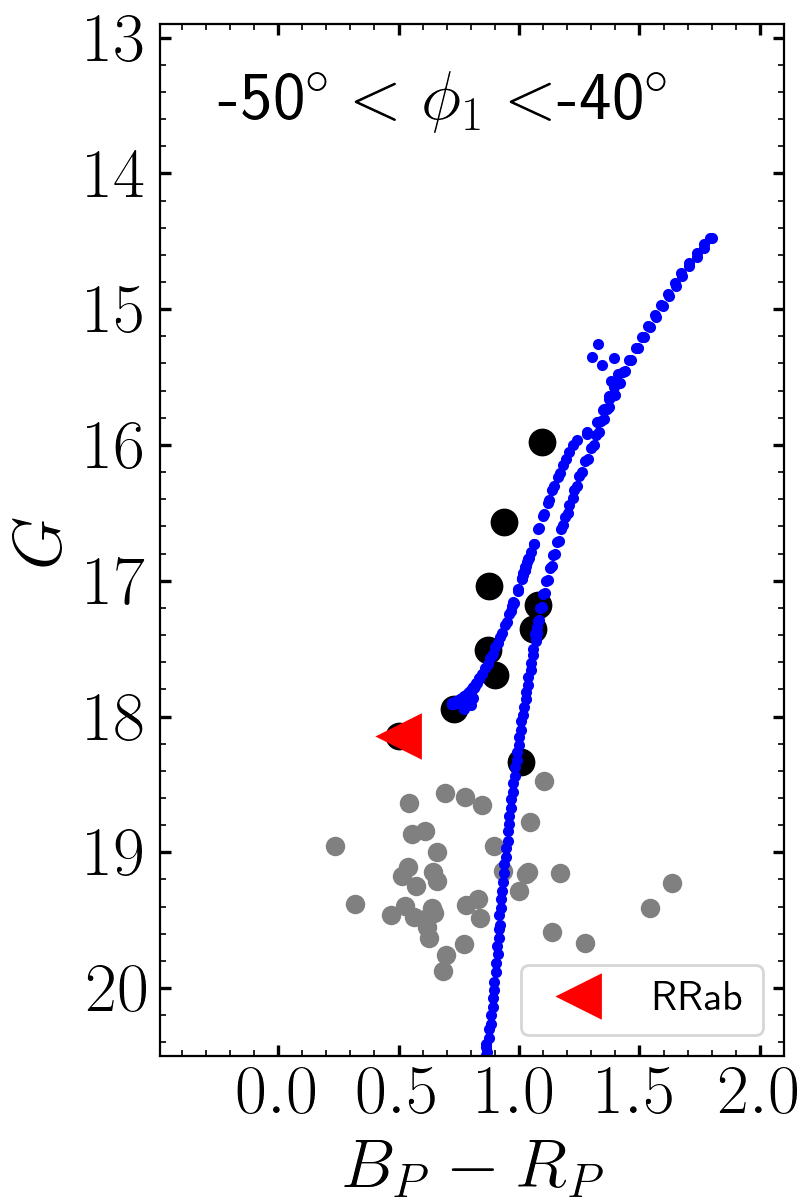} 
    \includegraphics[trim={1.9cm 0.5 0cm 0},clip,height=0.21\textwidth]{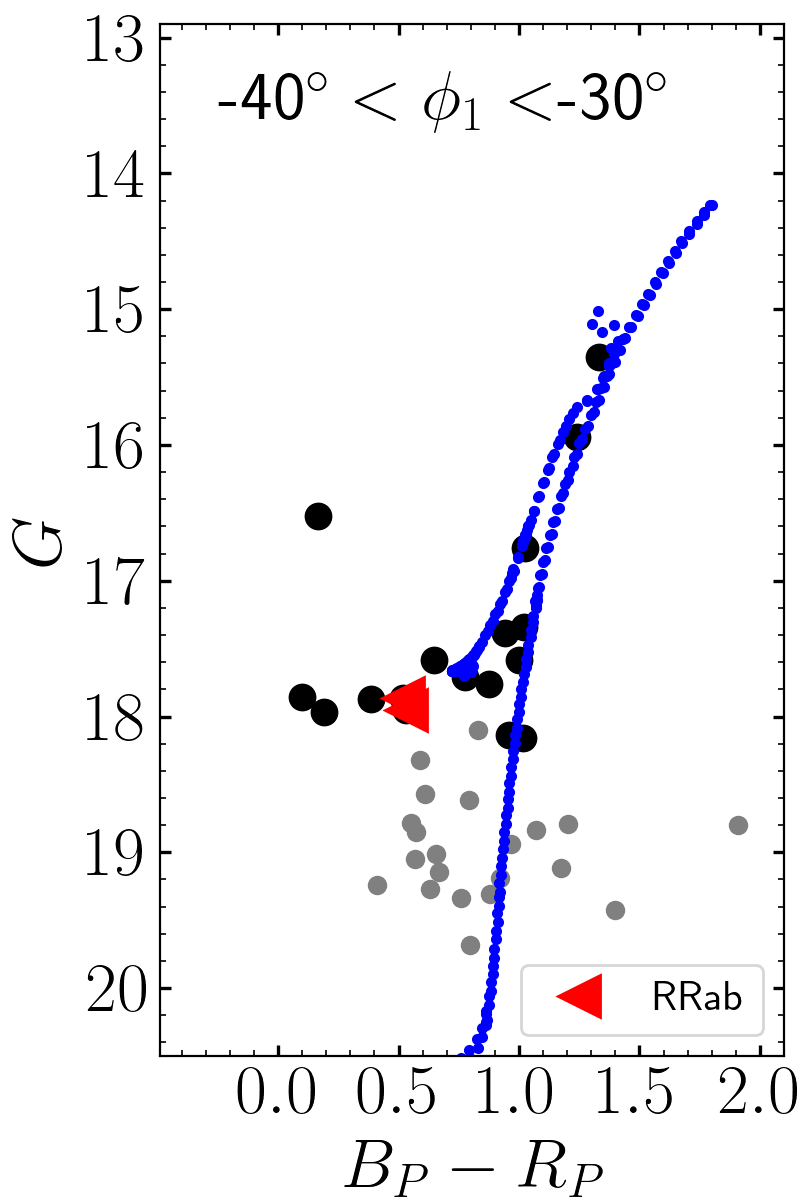} 
    \includegraphics[trim={1.9cm 0.5 0cm 0},clip,height=0.21\textwidth]{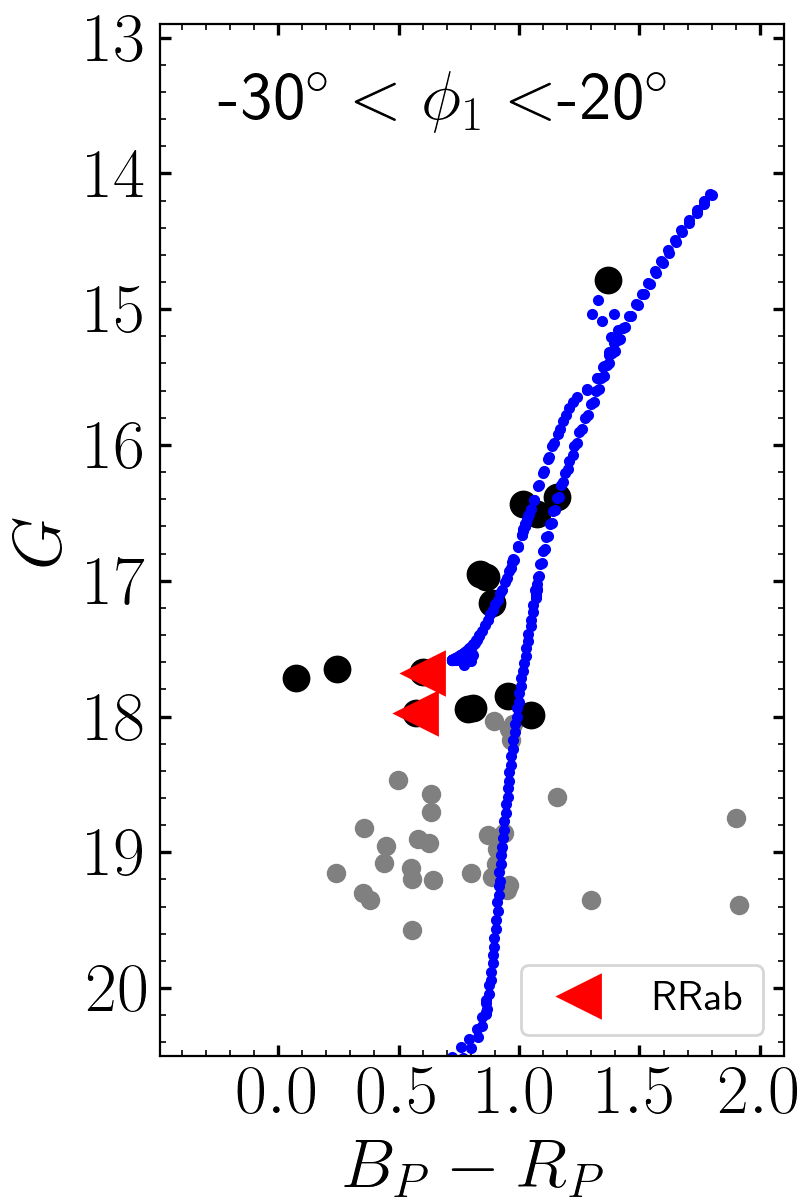} 
    \includegraphics[trim={1.9cm 0.5 0cm 0},clip,height=0.21\textwidth]{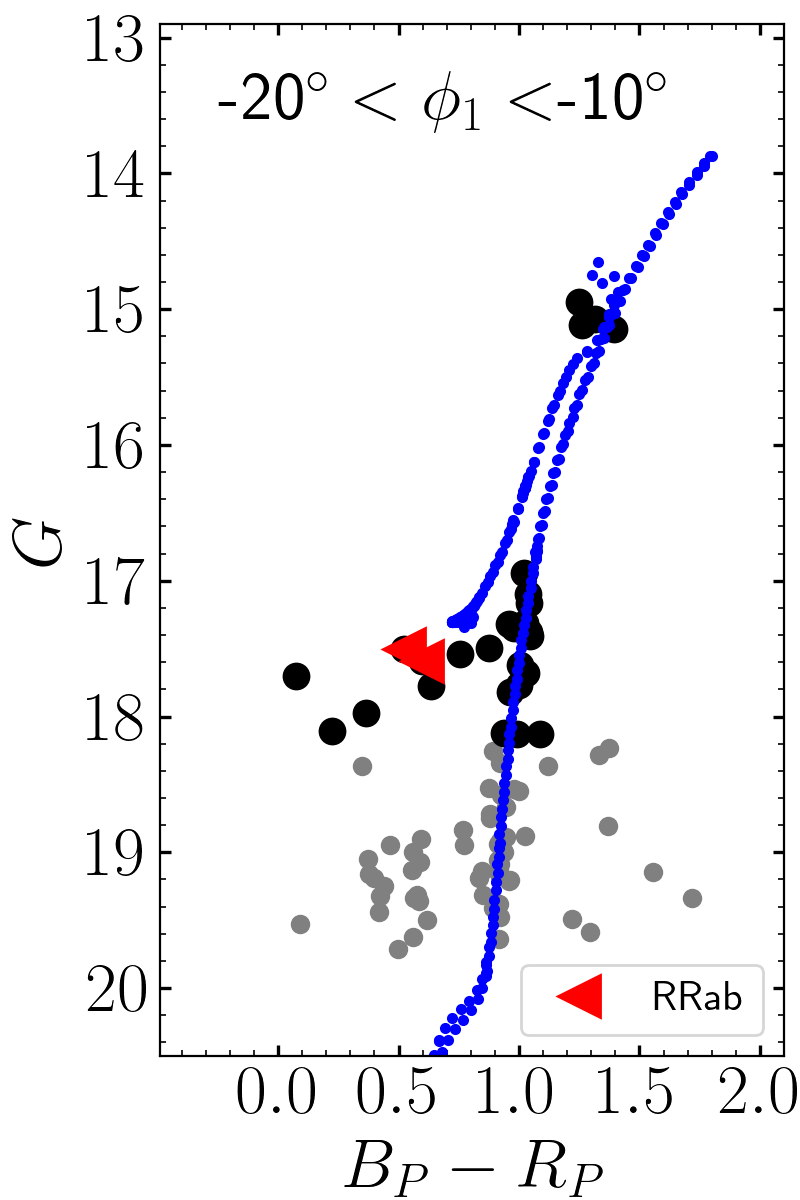}
    \includegraphics[trim={1.9cm 0.5 0cm 0},clip,height=0.21\textwidth]{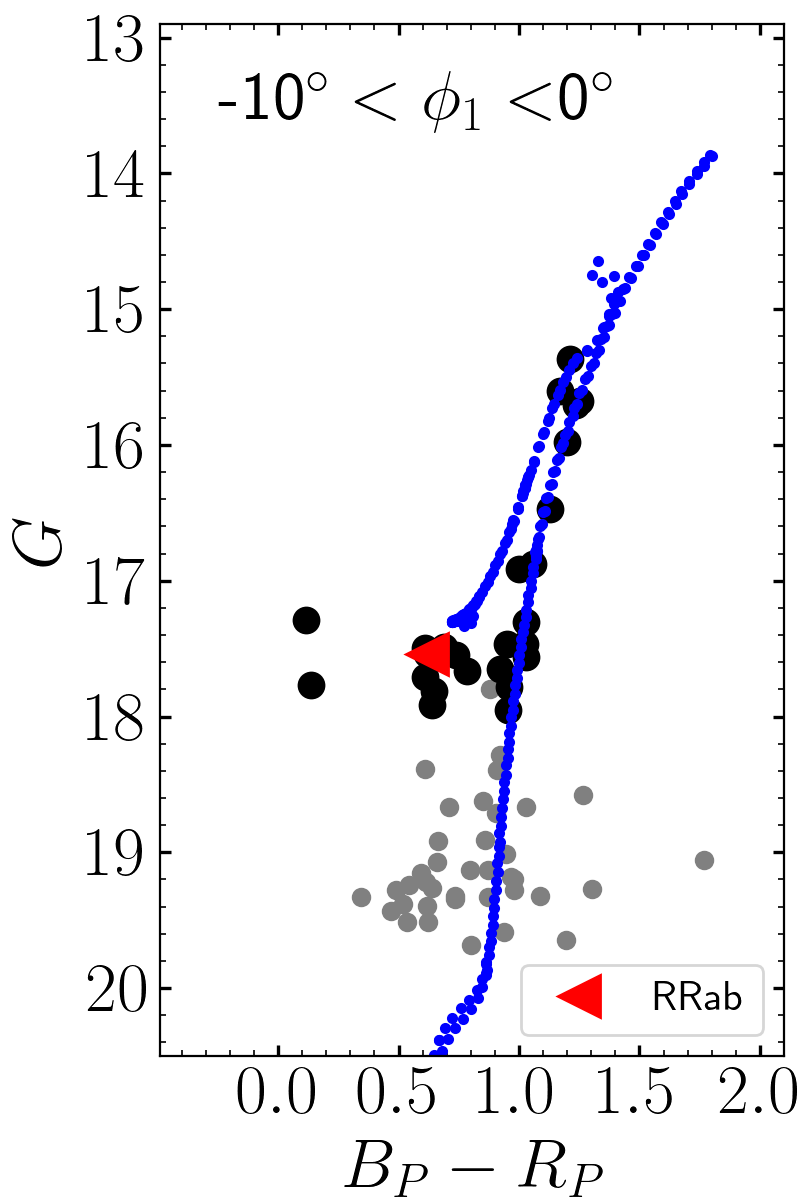} 
    \includegraphics[trim={1.9cm 0.5 0cm 0},clip,height=0.21\textwidth]{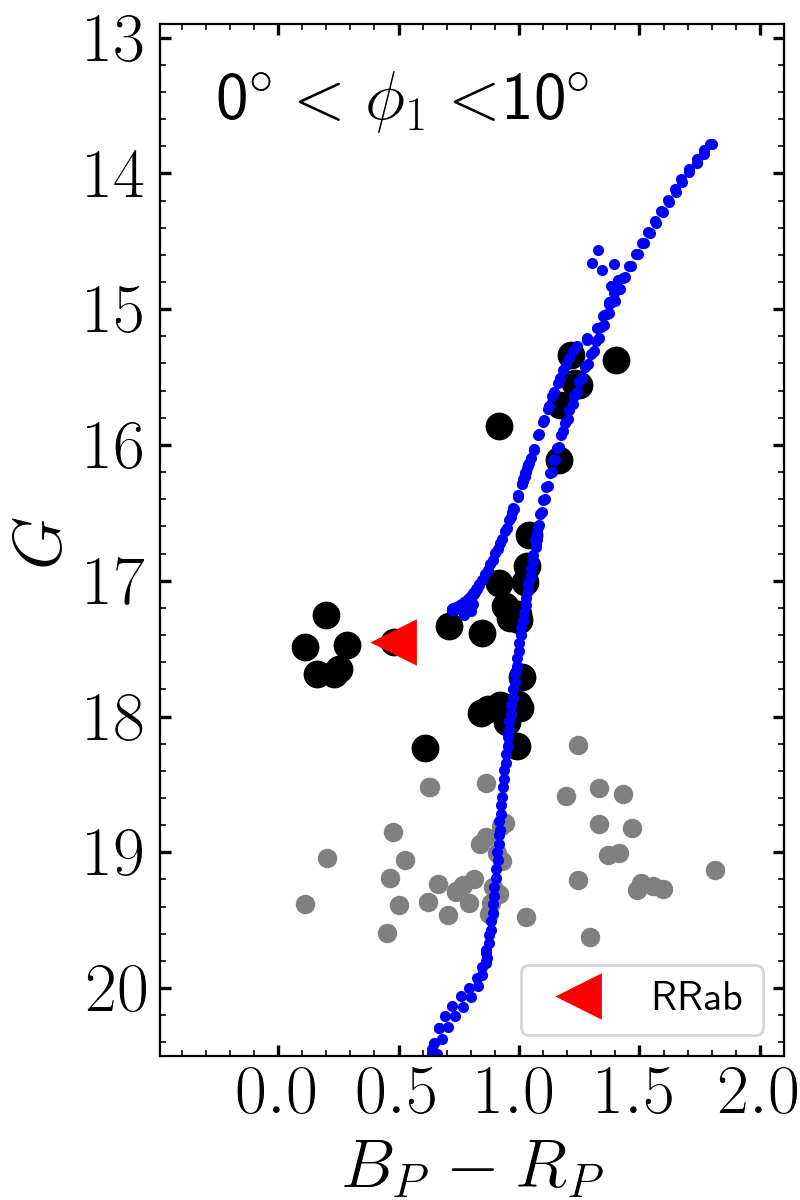} 
    \includegraphics[trim={1.9cm 0.5 0cm 0},clip,height=0.21\textwidth]{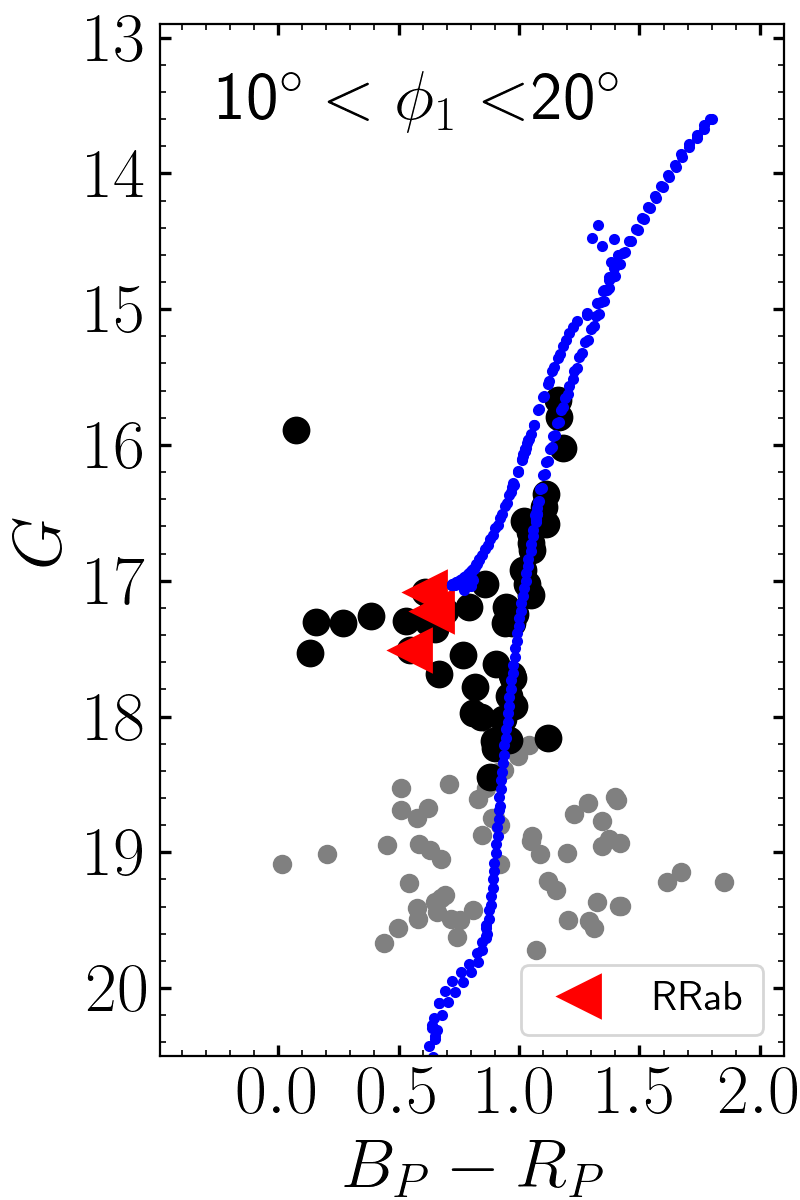} 
    \includegraphics[trim={1.9cm 0.5 0cm 0},clip,height=0.21\textwidth]{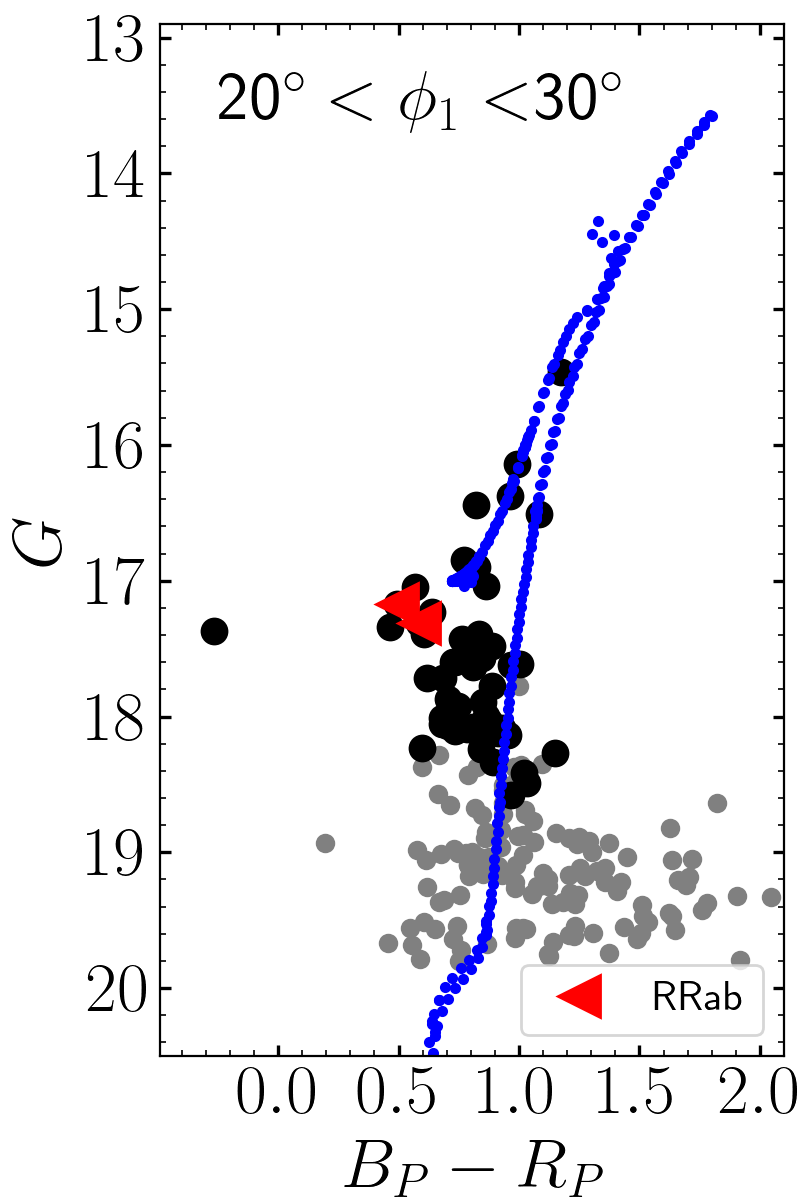} \\
    \vspace{0.2cm}

    {\bf CMD of control sample} \\ 
    \includegraphics[height=0.21\textwidth]{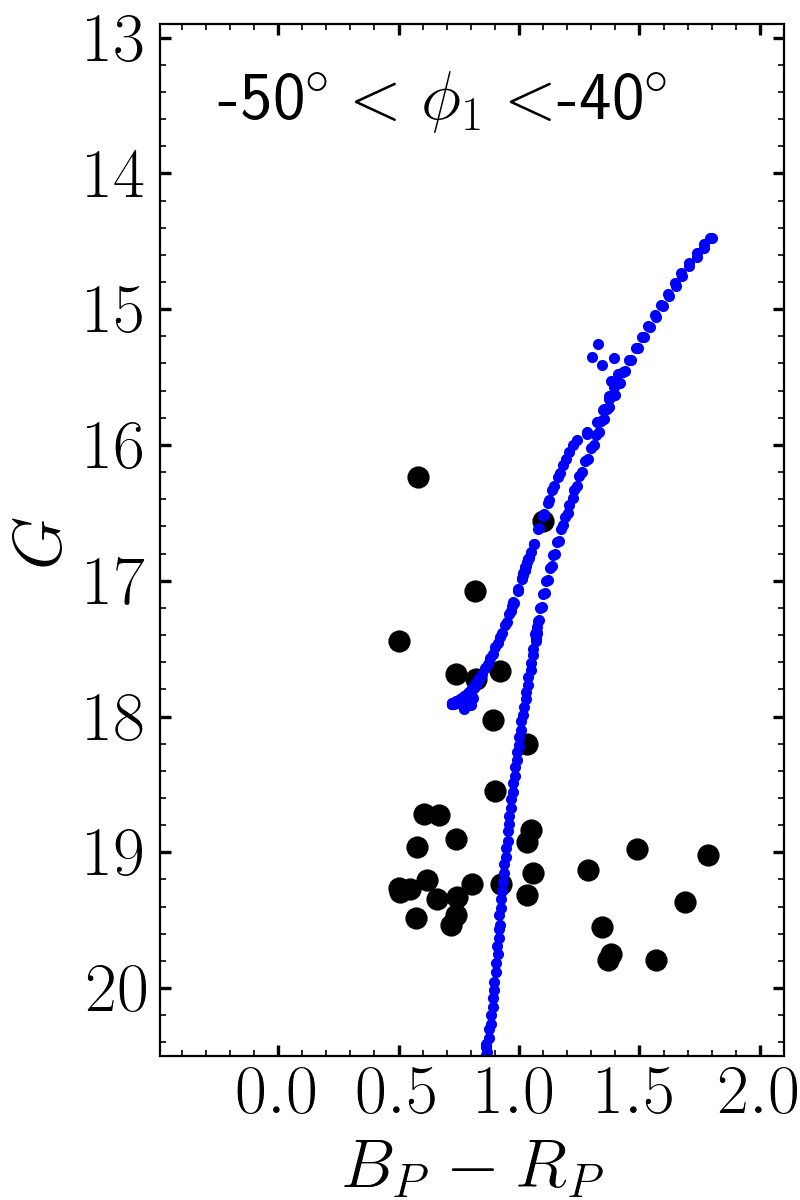} 
    \includegraphics[trim={1.9cm 0.5 0cm 0},clip,height=0.21\textwidth]{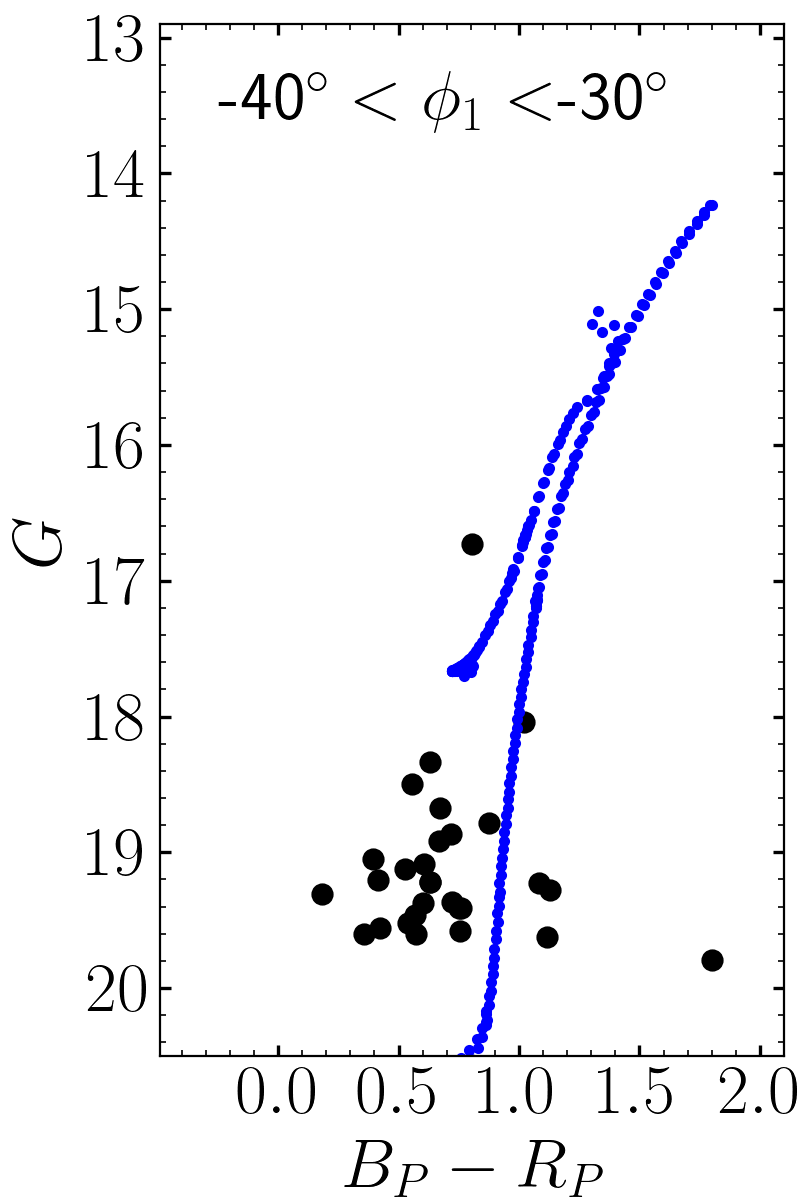} 
    \includegraphics[trim={1.9cm 0.5 0cm 0},clip,height=0.21\textwidth]{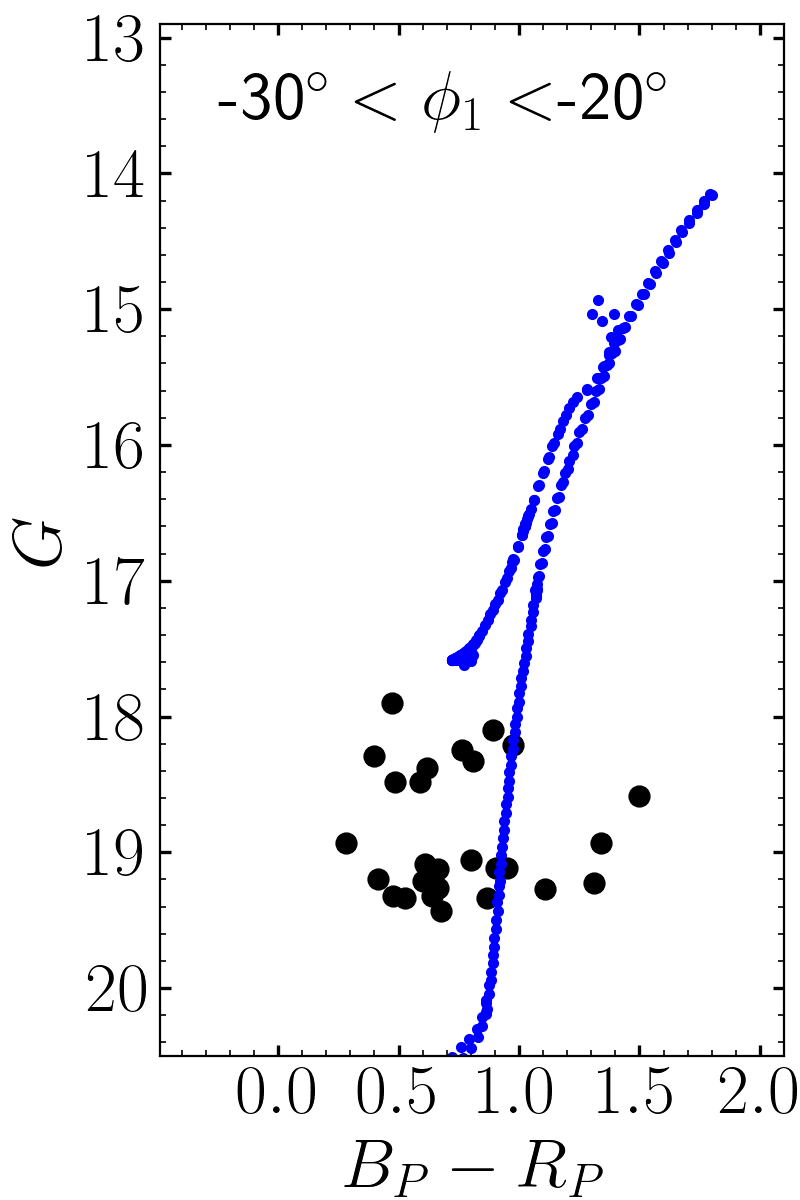} 
    \includegraphics[trim={1.9cm 0.5 0cm 0},clip,height=0.21\textwidth]{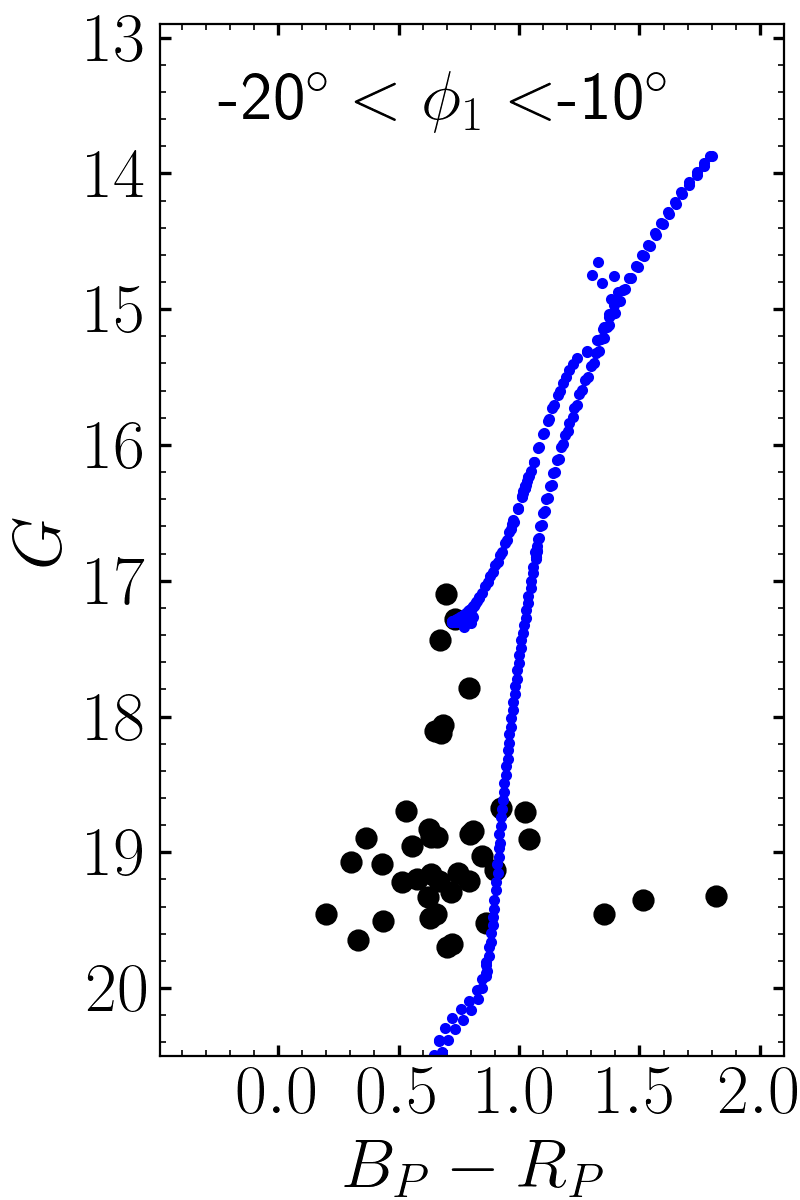}
    \includegraphics[trim={1.9cm 0.5 0cm 0},clip,height=0.21\textwidth]{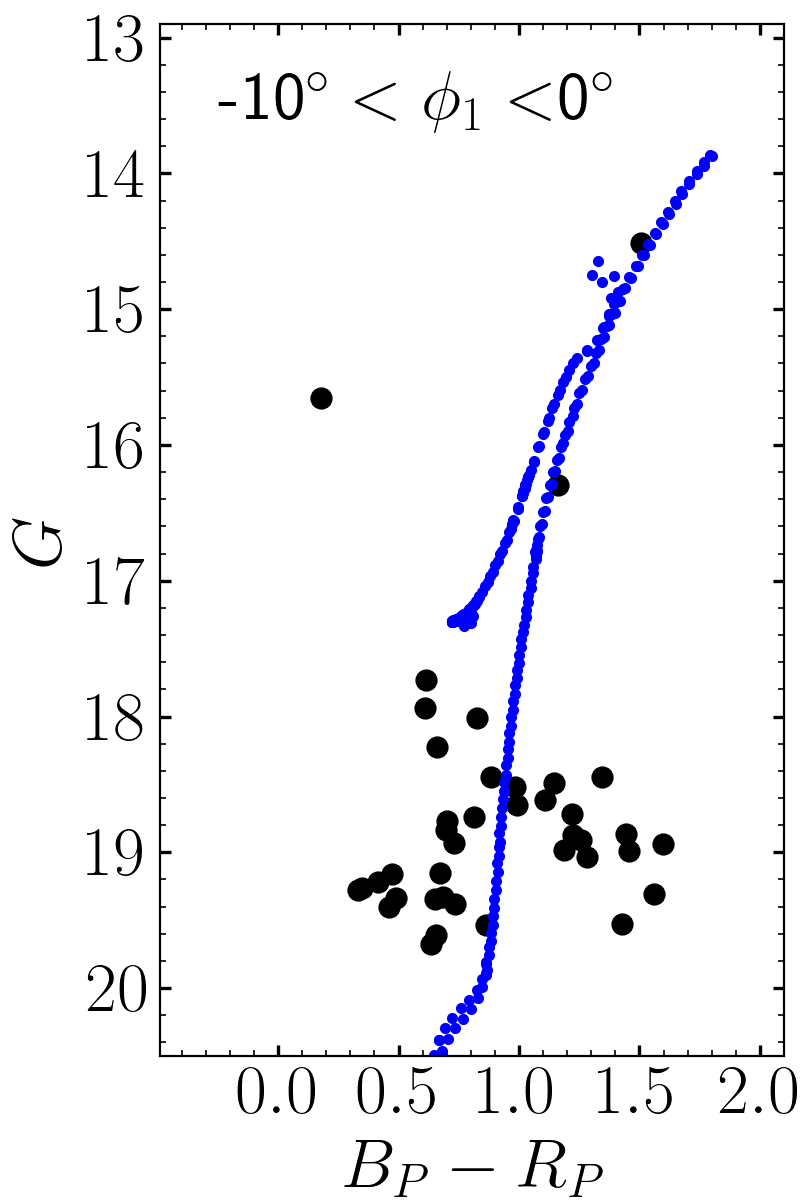} 
    \includegraphics[trim={1.9cm 0.5 0cm 0},clip,height=0.21\textwidth]{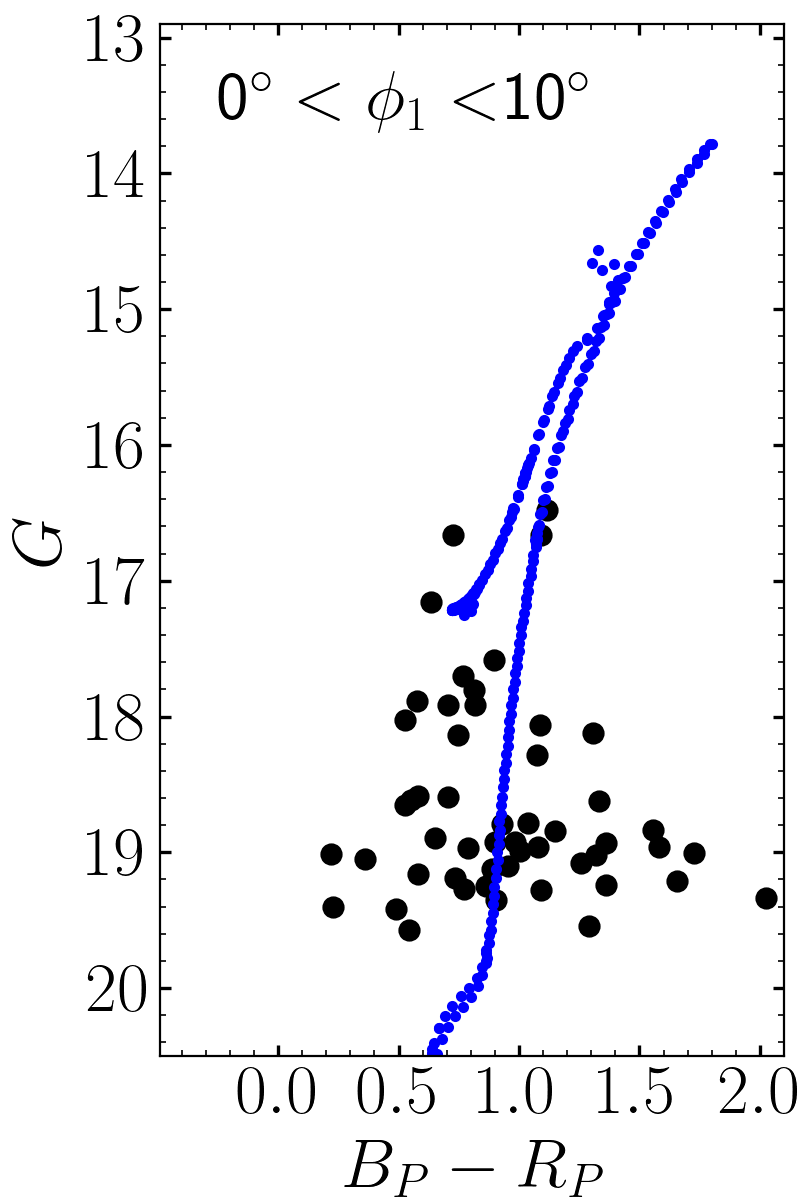} 
    \includegraphics[trim={1.9cm 0.5 0cm 0},clip,height=0.21\textwidth]{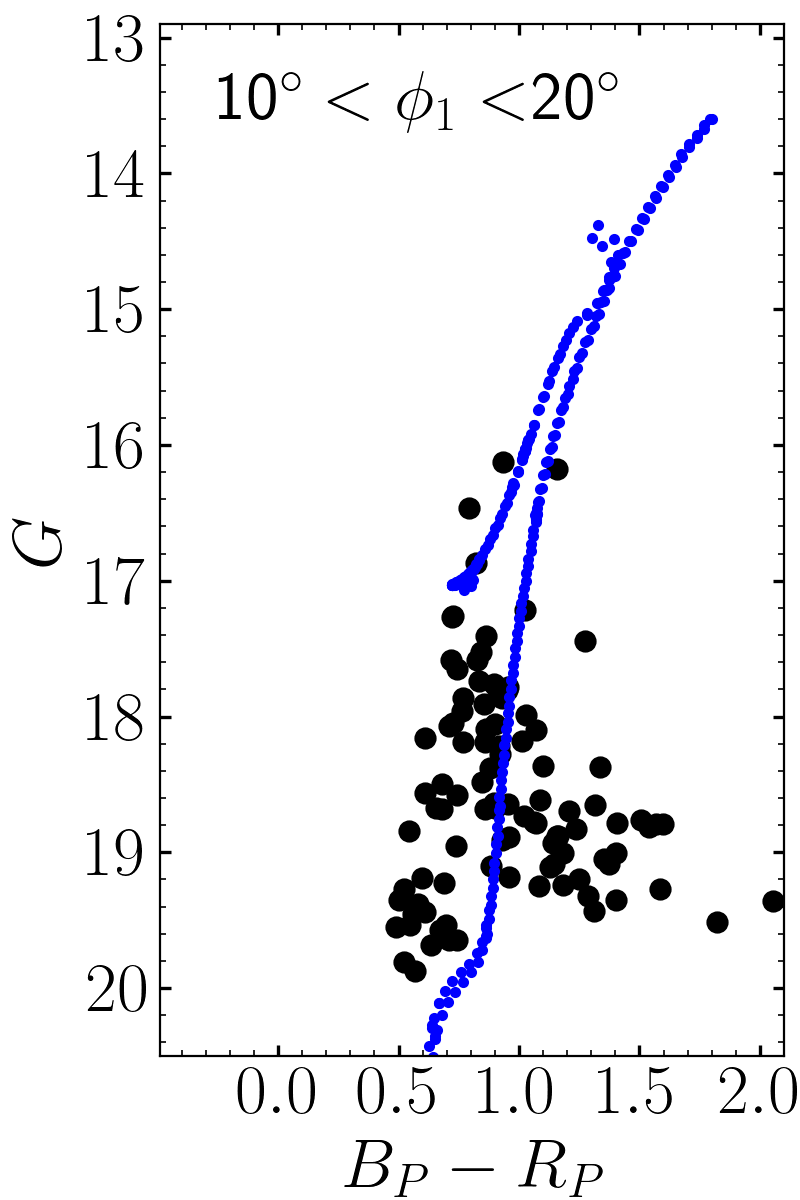} 
    \includegraphics[trim={1.9cm 0.5 0cm 0},clip,height=0.21\textwidth]{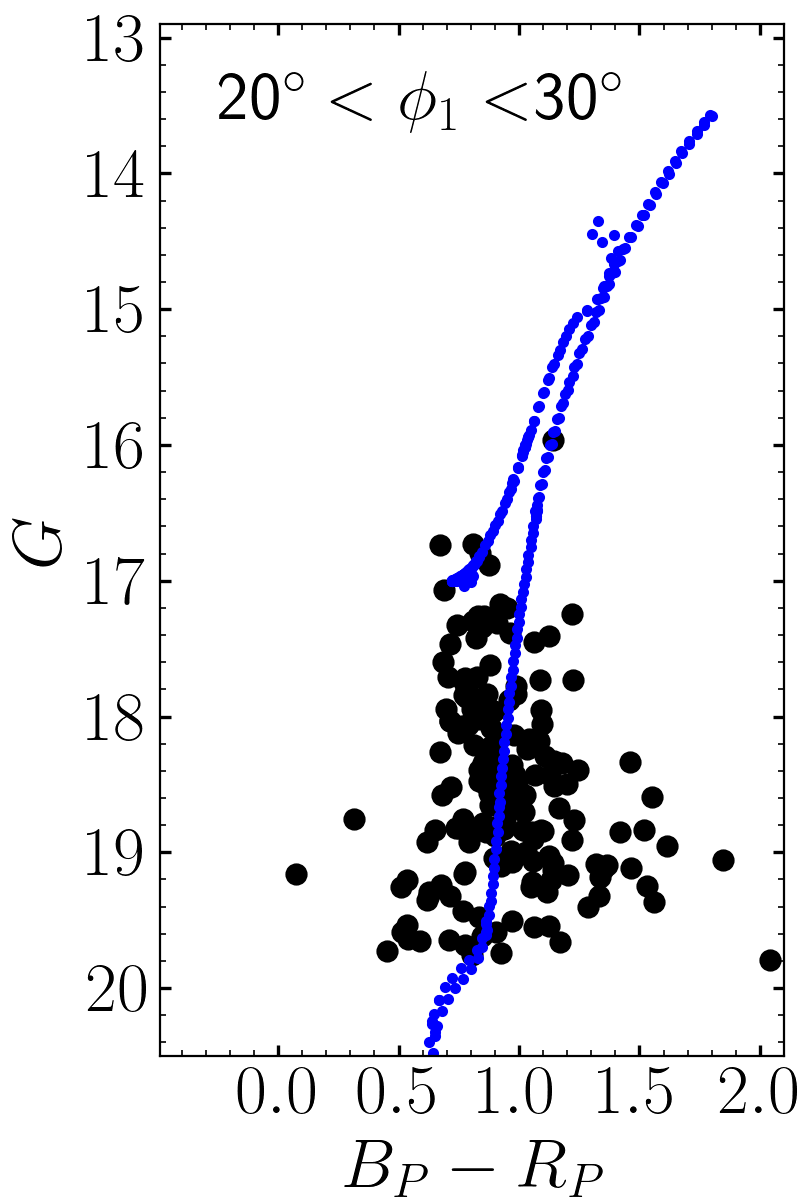} 
     \caption{The CMD distributions of the member candidate stars of the 
     stream in each subsample 
     are shown in the top panels. 
   The black and the gray dots
   are the stars with uncertainties  of proper motions smaller than $0.2$ and $0.5$ mas yr$^{-1}$, respectively.     The 
     isochrone with metallicity and age of $-1.3$ and $13$ Gyr is  
     represented by the blue dots with distance module derived from
     the RR Lyrae stars in the corresponding subsample, which are
     marked with red triangles. 
     Similar distributions of the control samples with $5^\circ<\phi_2<10^\circ$ and uncertainties smaller than $0.5$ mas yr$^{-1}$ are shown in the bottom panels.}
    \label{fig:cmd}
\end{figure*}

For further validation, we check the color-magnitude diagram (CMD) of the member candidates. 
Because of the large coverage on the sky,  we divide the member candidates
into subsamples according to 
the longitude $\phi_1$ with the step of  $10^\circ$. The distribution in the CMD of those candidates are represented 
by the black dots in  the top panels of Figure~\ref{fig:cmd}.  
Meanwhile, all the stars with proper motion uncertainties smaller
than $0.5$ mas yr$^{-1}$ are shown with gray dots. For comparison, control samples are selected with 
with $5^\circ<|\phi_2|<10^\circ$,  $-50^\circ<\phi_1<30^\circ$, and proper motion uncertainties smaller
than $0.5$ mas yr$^{-1}$. 
The CMD of those control samples are shown in the bottom panel of Figure~\ref{fig:cmd}.
In total, we obtained 8 member subsamples and 8 control samples. 

From the CMD of the member samples, we can find the significant signatures of 
the possible horizontal branch stars. 
To make sure if that is the horizontal branch, rather than the possible nearby turnoff stars, those fainter stars within 
the same proper motion ranges, but with larger uncertainties are also represented by the gray dots.
From the distribution of all the dots, we prefer that the those black dots around $G\sim18$ should be the 
horizontal branch stars.
 By cross-matching with the RR Lyrae catalog identified by \cite{2023A&A...674A..18C},
there are $14$ RR Lyrae stars (RRLs)  in the member candidates. 
We find that those RRLs are located with 
consistent position of the horizontal branch stars,
which indicates that all of those RRLs  should be the members of the 
stream. All  of those 14 RRLs are marked with red triangles in the corresponding panels.
Then, the distances of those RRLs  are calculated given a constant absolute magnitude in $G$-band, i.e. $0.63$ mag \citep{Muraveva2018MNRAS.481.1195M}. The 
resulting distances of these RRLs  are listed in  Table~\ref{tab:RRL}.

 \begin{figure}
    \centering
    \includegraphics[width=0.49\textwidth]{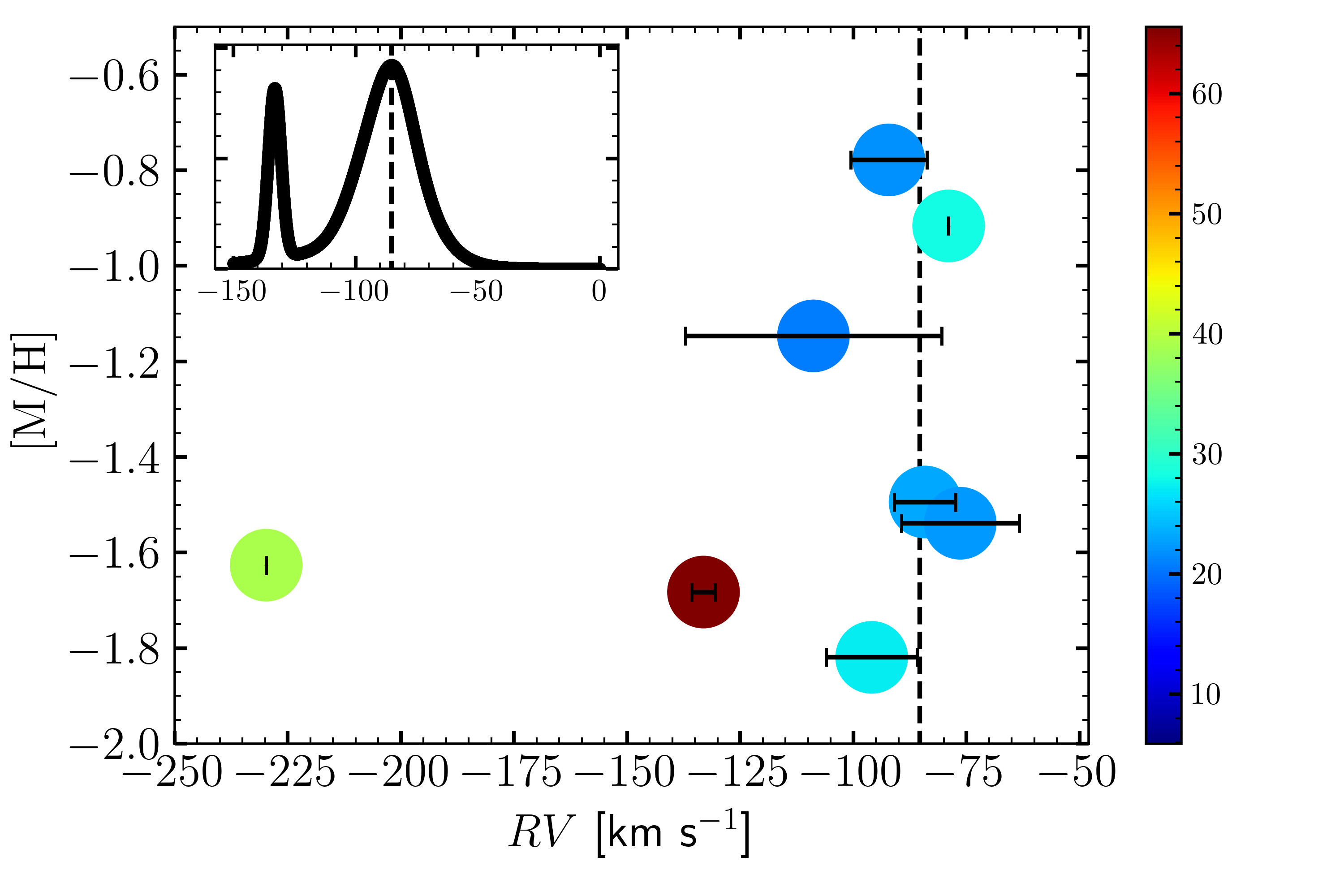} \\
     \caption{Eight  candidate stars observed by LAMOST are showed in the space of the radial velocity $RV$ 
     versus metallicity $\mathrm{[M/H]}$, color coded by the signal-to-noise ratio in $i-$band. The subplot
     shows the probability distribution of the radial velocities evolving 
     Gaussian kernel with width of the uncertainties of $RV$.
     The median uncertainty $8.4$ km s$^{-1}$ is adopted for the two stars 
     whose uncertainties are not given.}
    \label{fig:MH_RV}
\end{figure}
\begin{figure}
    \centering
    
    \includegraphics[width=0.48\textwidth]{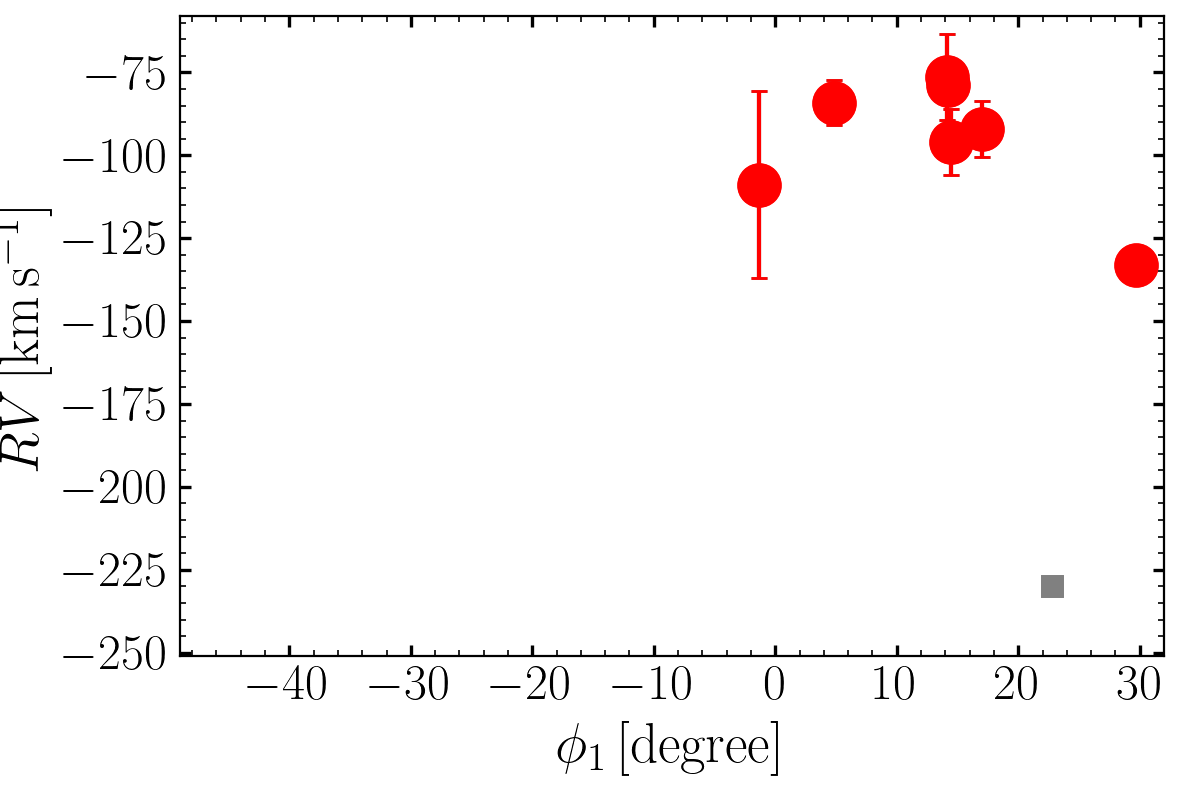} \\ \vspace{-1.1cm}
    \includegraphics[width=0.48\textwidth]{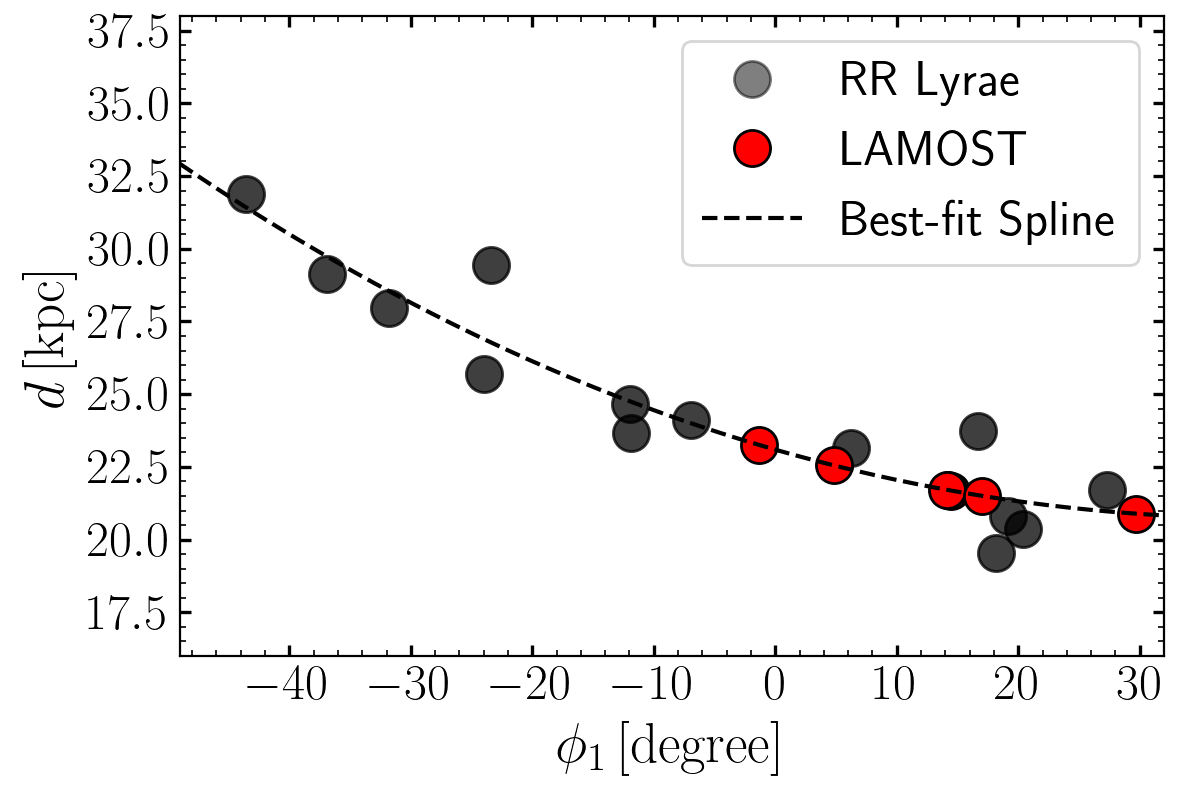} \\ \vspace{-1.1cm}
    \includegraphics[width=0.48\textwidth]{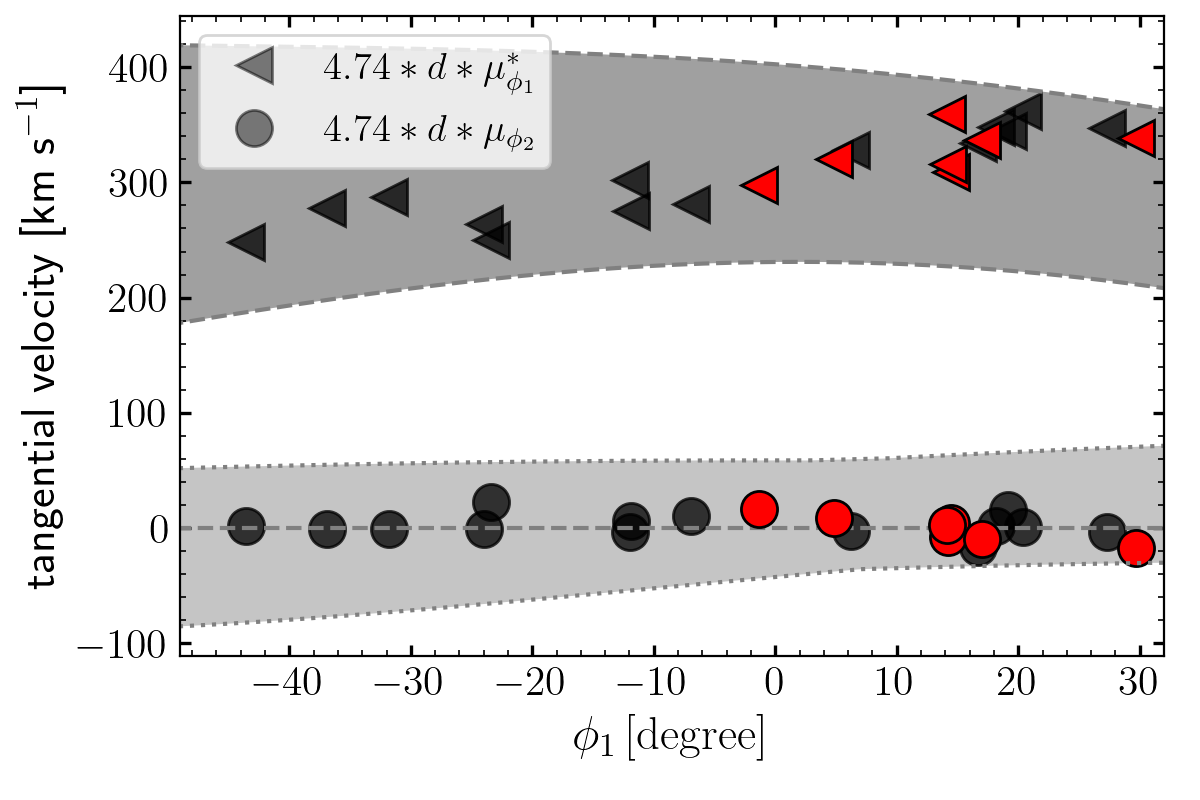} \\
    
     \caption{ The variances of the radial velocity $RV$, the distance $d$ and tangential velocities of the RRLs (black dots) are shown 
     as a function of longitude $\phi_1$ from the top to the  bottom panels, respectively. 
      $7$ of the LAMOST member stars are represented by the red dots.
     The dashed line in the middle panel 
     represents the best-fit spline.  In the bottom panel,
     the filled black triangles and dots show the two tangential 
     velocity components along $\phi_1$ and $\phi_2$ of the RRLs, respectively. 
     The dark and light gray regions represent the limits
     for the two tangential velocities, which are caused 
by the proper motion cut during sample collection.
     The red dots represent the tangential velocity distributions
     of the  7 LAMOST member stars.
     The horizontal dashed line in the bottom panel indicates the zero velocity along $\phi_2$. }
    \label{fig:dvv}
\end{figure}

Except the RRLs, there are another $8$ stars in the member candidates, 
which are included in LAMOST DR9 \citep{LAMOST2020arXiv200507210L}. 
The radial velocity ($RV$) and the metallicity $\mathrm{[M/H]}$
have been estimated by SLAM \citep{ZhangBo2020ApJS..246....9Z}. Figure~\ref{fig:MH_RV} 
shows the distribution in [M/H] vs. $RV$ plane of the $8$ stars with color-coded signal-to-noise
ratio in $i-$band (SNR$_i$).
With a Gaussian kernel involving the uncertainties of $RV$, 
we obtain the kernel-smoothed distribution of $RV$. There is a clear
peak showing at $RV\sim-87.2$ km s$^{-1}$ in the $RV$ distribution 
displayed in the subplot in the top left inset. This peak represents 
the characteristic radial velocity of the stream. It is noted that there are two stars
whose uncertainties are not provided. A median uncertainty of $8$ km s$^{-1}$ 
is then adopted for them during the calculation of the kernel-smoothed distribution. 
 Then all $7$ of the $8$ stars with $RV>-150$ km s$^{-1}$ are confirmed as the members
(hereafter LAMOST member stars), including
one horizontal branch star and six red giant branch stars. 
The radial velocity distribution versus the longitude $\phi_1$ of the $8$ stars  is shown in the top panel
of Figure~\ref{fig:dvv}. All those selected $7$ stars 
are marked with red dots.
The mean metallicity $\mathrm{[M/H]}=-1.3$ of them may represent the metallicity of the stream. 

We then adopt the isochrone with metallicity $\mathrm{[M/H]}=-1.3$ 
and age $\tau=12$ Gyr to fit the stream member candidates, which are represented 
by the blue dots in Figure~\ref{fig:cmd}. The isochrone is shifted according to 
the distances of the RRLs in the corresponding panel. Note that we do not find RRLs
in the control samples.  The isochrone in each subsample is shifted with the same
distance as the member sample with same $\phi_1$ range.


\section{Properties of the stream}
With those 14 RRLs identified by Gaia and the $7$ LAMOST member stars with radial 
velocities, we can study the geometric and kinematic features of the stream.

\subsection{The spatial distribution}

 \begin{figure*}
    \centering
    \includegraphics[height=0.25\textwidth]{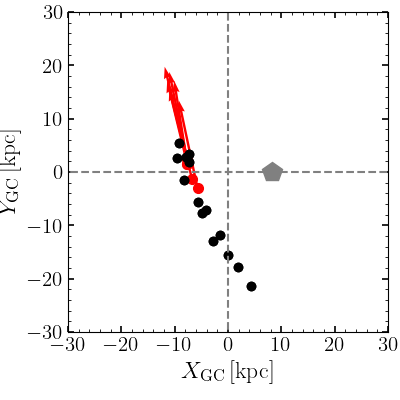} 
    \includegraphics[height=0.25\textwidth]{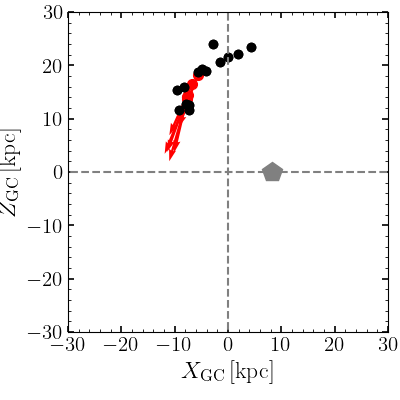} \hspace{-0.2cm}
    \includegraphics[trim={1.6cm 0 0cm 0},clip,height=0.25\textwidth]{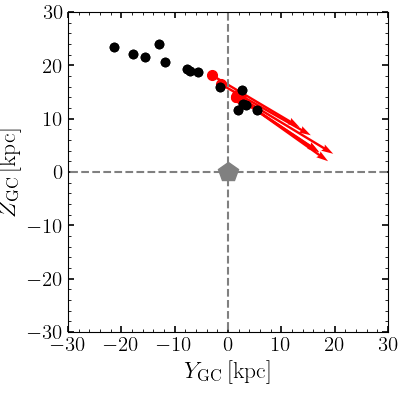} \hspace{-0.2cm}
    \includegraphics[trim={1.6cm 0 0cm 0},clip,height=0.25\textwidth]{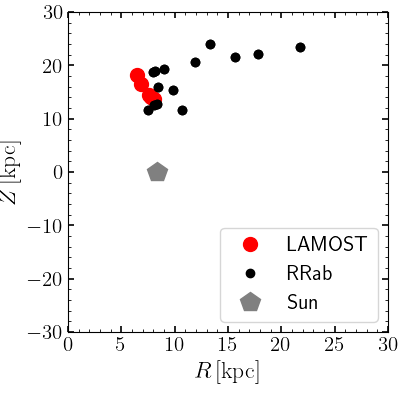} \hspace{-0.2cm} \\ 
     \caption{The 14 RRLs and {\bf $7$ LAMOST} member stars of the stream are represented in the spaces
     with black and red dots, respectively. 
     The absolute magnitude 0.63  \citep{Muraveva2018MNRAS.481.1195M} is adopted for distance 
     calculation of the RRLs.The movements of the {\bf $7$ LAMOST member stars} are
     represented by the red arrows. The position of the Sun is marked as the gray
     pentagon. The dashed lines represent the position of the Galactic center.}
    \label{fig:space}
\end{figure*}
First, the distance distribution of the 14 RRLs is shown in the middle panel of 
Figure~\ref{fig:dvv}. 
 We fit the distance as a function of $\phi_1$ with a spline function
with a large smoothing factor $s=5000$, i.e. \textit{UnivariateSpline} from 
the package \texttt{scipy}. The best-fit line is represented by the 
dashed line in the middle panel of Figure~\ref{fig:dvv}. 

Then, the distances of the  $7$ LAMOST member  stars are 
estimated by applying the above relation. Note that those  $7$  stars are located with
$\phi_1>-10^\circ$
because LAMOST can only observe stars with $\delta>-10^\circ$, 
the whole stream cannot be traced completely with LAMOST data. 
Figure~\ref{fig:space} shows the distributions in the Cartesian $X$-$Y$ plane of the 
$14$ RRLs and those  $7$ LAMOST member stars with black and red dots, respectively. 
The Galactocentric distance $r$ of the RRLs varies from $12.61$ to $33.57$ kpc, 
which provides a lower limit of the eccentricity
\begin{equation}
e_{\mathrm{lower}}=\frac{r_{GC}^{max}-r_{GC}^{min}}{r_{GC}^{max}+r_{GC}^{min}}=0.45
\end{equation}
which means that the stream has a radial orbit.

 \begin{figure}
    \centering
    \includegraphics[width=0.49\textwidth]{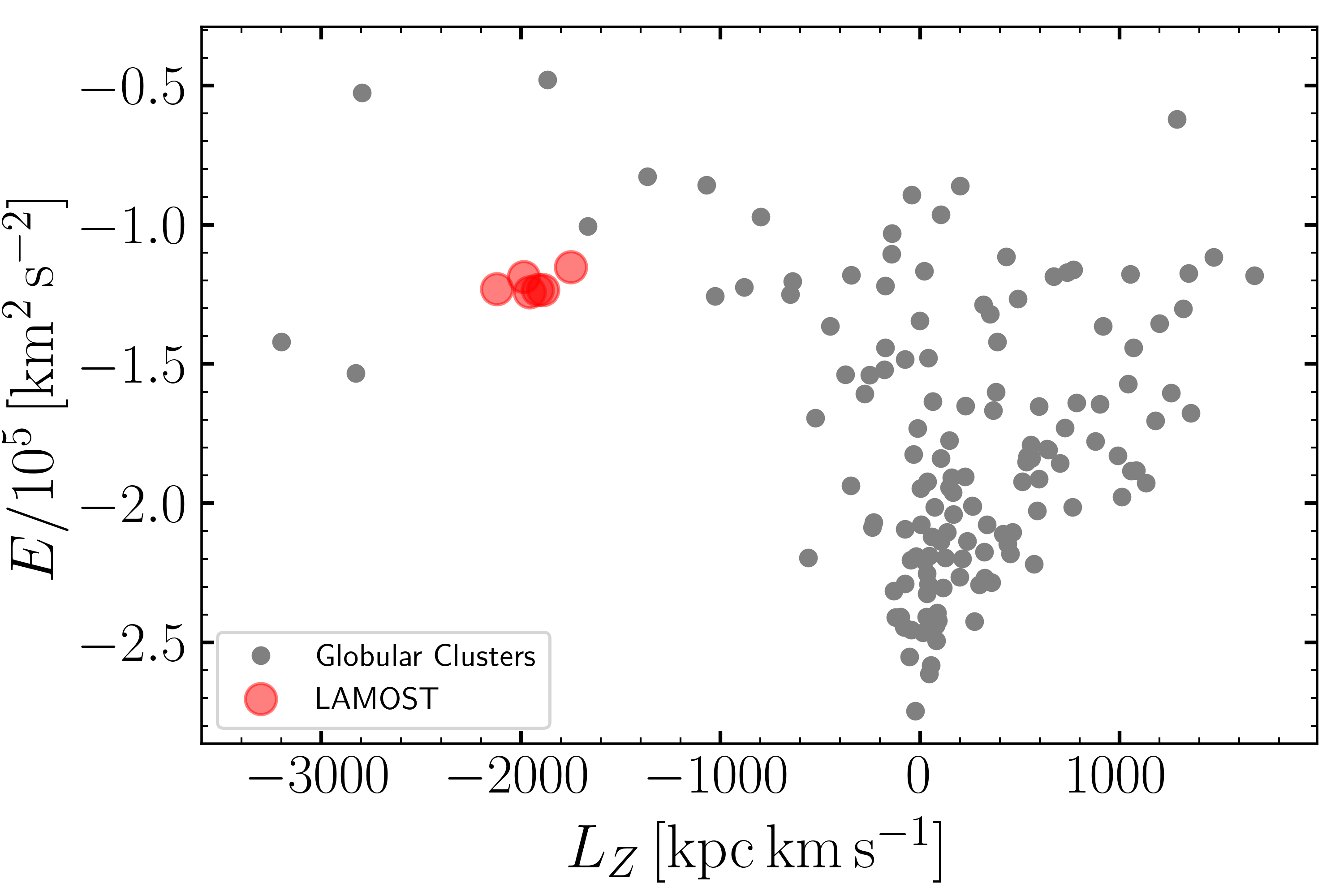}
     \caption{The {\bf $7$ LAMOST member stars} and the globular clusters \citep{Vasiliev2021MNRAS.505.5978V} are represented in 
     the phase space $E$ versus $L_Z$ with red dots and gray dots, respectively. }
    \label{fig:E_Lz}
\end{figure}

\subsection{Orbit property}
Currently, we do not have the radial velocities for all the member candidates of the stream.
However, the RRLs can provide  information about the orbit of the stream, because of their 
accurate distance estimation. The proper motions can be converted to the tangential 
velocities. We, therefore, remove the contribution of the Solar motion,
i.e. $(U,\, V,\, W)=(11.1, \,12.24+232,\, 7.25)$ km s$^{-1}$ from the observed 
proper motion. The proper motions with respect to the position of the Sun 
are then converted to the coordinates of the stream with \texttt{gala} again, i.e. $(\mu_{\phi_1^*}^C,\mu_{\phi_2}^C)$.
Finally, the tangential velocities of the RRL candidates w.r.t. 
the position of the Sun are calculated by $V_{\phi_1^*}=4.74\,d\,\mu_{\phi_1^*}$ and
$V_{\phi_2}=4.74\,d\,\mu_{\phi_2}$. The two tangential velocity components are 
displayed in the bottom panel of Figure~\ref{fig:dvv}. 
The dark and light gray regions represent the limits which are caused 
by the proper motion cut during sample collection. The upper and lower 
limits are estimated as follows. Firstly we generate  
10000 stars at each longitude $\phi_1$ (from $-50^\circ$ to $30^\circ$ with step $0.45^\circ$), 
with proper motions and latitude $\phi_2$ uniformly distributed in the 
corresponding ranges discussed above, i.e.  $-1.0<\mu_\delta<-0.5$ mas yr$^{-1}$, 
$0.5<\mu_\alpha^*<2.0$ mas yr$^{-1}$ and $-5^\circ<\phi_2<5^\circ$.
Then we repeat  the same correction steps to remove the contribution of the 
Solar motion. The maximum and minimum values of the two tangential 
velocities are represented 
by the upper and lower edges of both the shadow regions.
It shows that $V_{\phi_1^*}$ 
smoothly increases as a function of $\phi_1$ and $V_{\phi_2}$ is strictly along zero
with small random dispersion. Since that the increasing direction of $\phi_1$ is 
approaching the Galactic center and the distance to the Galactic center declines 
with increasing $\phi_1$, the increasing $V_{\phi_1^*}$ implies that the member 
stars of the stream accelerate when they are approaching the pericenter point 
close to the Galactic center. 
Meanwhile, the essential zero value of $V_{\phi_2}$ of the RRLs is evident 
that they move exactly along the stream. 

The motions of the RRLs are also represented with arrows color-coded by their distances in 
Figure~\ref{fig:radec}. Similar to what we have seen in the bottom panel of 
Figure~\ref{fig:dvv}, the stream is perfectly moving along the stream and 
approaching  the Galactic center. 

The distances of the  $7$ LAMOST member stars are estimated by using the $d-\phi_1$ relation fitted by 
the spline (the dashed line in the top panel of Figure~\ref{fig:dvv}) from the RRLs, 
which are represented by the red symbols in Figure~\ref{fig:dvv}. 
Then, their 3D velocities can be derived and are represented by the red arrows in
Figure~\ref{fig:space}. Like the tangential velocities of the RRLs,
it also shows that the $6$ stars move along the stream. 
With the full 6D information of those LAMOST stars, we estimate their total energy and the angular momenta with \texttt{AGAMA} \citep{Vasiliev2019MNRAS.482.1525V} 
 with a potential including a Dehnen bulge of mass $2\times10^7M_\odot$ and scale radius $1.0$ kpc, 
a Miyamoto-Nagai disk of $5\times10^{10}M_\odot$, scale radius $3.0$ kpc and scale height $0.3$ kpc, 
and an NFW halo with mass of $5.5\times10^{11}M_\odot$ and scale radius $15.0$ kpc.
As shown in Figure~\ref{fig:E_Lz}, the LAMOST member stars 
show a clear retrograde rotation. Meanwhile, the average 
eccentricity of those 6 member stars is around $0.58$, which is consistent 
with the coarse estimate from the spatial distributions of the RRLs. 
As a comparison, the globular clusters are also represented as gray dots \citep{Vasiliev2021MNRAS.505.5978V}. 
The large total energy $E$ and the retrograde motion indicate that this stream 
is possibly associated with the merging event
{\it Sequoia}. This discovery indicates that there should be more streams in the halo,
which are not  discovered yet because of the observational and methodological limits, especially for those streams
with highly radial orbits and low surface luminosity.

\section{Summary}
With accurate astrometry measurements, 
we select a catalog of distant stars with accurate proper motions $\sigma_{\mu_\alpha^*}<0.2$ and $\sigma_{\mu_\delta}<0.2$ mas yr$^{-1}$. From the density distributions with different proper motion ranges, 
we find a new stellar stream, which is $\sim80^\circ$ long and $\sim1.70^\circ$ wide 
($1\sigma_{\phi_2}$ assuming a Gaussian distribution). According to the spectra of  $7$ LAMOST 
member stars, the stream has a metallicity
of $\mathrm{[M/H]}=-1.3$, which is  consistent with that of the merging event {\it Sequoia} \citep{Feuillet2021MNRAS.508.1489F}.
The location of the stream in  phase space $E$ versus $L_Z$ also indicates the same association.

With all $14$ RR Lyrae stars and $7$ LAMOST member stars, we find that the stream has an eccentric orbit with $e\sim0.58$. 
 Assuming the stream has at least 14 RRLs, its progenitor should be of
luminosity brighter than  $M_V=-6.9$ \citep{Mateu2018MNRAS.474.4112M}.
To further study the substructure, spectroscopic observations are 
necessary to constrain their chemical and full dynamical features, such as 
LAMOST \citep{LAMOST2020arXiv200507210L}, 
SDSS-V\footnote{https://www.sdss.org/},  
DESI \citep{DESI2023ApJ...947...37C}, 
S$^5$ \citep{S52019MNRAS.490.3508L}, 
H3 \citep{H32019ApJ...883..107C}, 
and future WEAVE\footnote{https://www.ing.iac.es/weave/}, 4MOST\footnote{https://www.4most.eu/cms/home/} and, PFS \citep{PFS2014PASJ...66R...1T} surveys.

\begin{acknowledgments}
We thank Dongwei Fan from NAOC for the help on using the Chinese Virtual Observatory.
This work is supported by National Key R\&D Program of 
China No. 2019YFA0405500.
H.T. is supported by National Natural Science Foundation of China with grant No. 12103062,12173046, U2031143.
and Beijing Natural Science Foundation with grant No. 1232032.
C.L. thanks the science research grants from the China Manned Space Project with NO. CMS-CSST-2021-A08.
Y.Y. acknowledge the support from National Natural Science Foundation of China with Grant No.12203064.
X-X.X. acknowledge the support from CAS Project for Young Scientists in Basic Research Grant 
No. YSBR-062 and NSFC grants No. 11988101, 11890694. 
We acknowledge the science research grants from the China Manned Space Project with NO. CMS-CSST-2021-B03.
Data resources are supported by China National Astronomical Data Center (NADC) and Chinese Virtual Observatory (China-VO). 
This work is supported by Astronomical Big Data Joint Research Center, co-founded by National Astronomical Observatories, 
Chinese Academy of Sciences and Alibaba Cloud.

This work has made use of data from the European Space Agency (ESA)
mission {\it Gaia} (\url{https://www.cosmos.esa.int/gaia}), processed by
the {\it Gaia} Data Processing and Analysis Consortium (DPAC,
\url{https://www.cosmos.esa.int/web/gaia/dpac/consortium}). Funding
for the DPAC has been provided by national institutions, in particular
the institutions participating in the {\it Gaia} Multilateral Agreement.
\end{acknowledgments}

%

\vspace{5mm}


\software{\texttt{Healpy} \citep{healpy_2005ApJ...622..759G,healpy_Zonca2019},
		\texttt{NumPy} \citep{harris2020array},
		 \texttt{Matplotlib} \citep{Hunter2007}
          }



\appendix

\section{Information of the RR Lyrae stars}
The information of the member RR Lyrae stars of the stream is listed in the Table~\ref{tab:RRL}.
\begin{rotatetable}

\begin{deluxetable*}{rrrrrrrrrrrrr}
\tabletypesize{\tiny}
\tablecaption{The member RR Lyrae stars of the  new stream are listed.
     Columns from left to the right are the source ID from Gaia DR3, the coordinates ($\alpha$, $\delta$),  
     the proper motions, the radial velocities and its uncertainty, the signal-to-noise ratio and the distance 
     obtained from the interpolation with RRL stars.
    corrected.}\label{tab:RRL}
    \tablehead{
  \colhead{source ID} &\colhead{$\alpha$} & \colhead{$\delta$} & \colhead{$d$}  & \colhead{$G$}   & \colhead{$\mu_\alpha^*$} & \colhead{$\mu_\delta$}  & \colhead{$\sigma_{\mu_\alpha^*}$} & \colhead{$\sigma_{\mu_\delta}$} & \colhead{$\phi_1$} & \colhead{$\phi_2$} & \colhead{$\mu_{\phi_1^*}^C$} &  \colhead{$\mu_{\phi_2}^C$}    \\   
            &\colhead{$^\circ$}   & \colhead{$^\circ$}    & \colhead{kpc}   & \colhead{mag}    & \colhead{mas yr$^{-1}$}& \colhead{mas yr$^{-1}$}  & \colhead{mas yr$^{-1}$}   & \colhead{mas yr$^{-1}$} & \colhead{$^\circ$}     & \colhead{$^\circ$}     & \colhead{mas yr$^{-1}$}& \colhead{mas yr$^{-1}$}} 
\startdata
 3573282914757958144 & $177.5353$ & $-13.1198$  & $33.59$ & $18.26$  & $0.862$  &  $-0.716$   & $0.157$  & $0.102$  & $-43.5995$   &  $+1.0053$   & $1.597$  & $-0.030$   \\
 3577098696846190976 & $184.3689$ & $-12.4451$  & $30.97$ & $18.11$  & $1.055$  &  $-0.753$   & $0.162$  & $0.149$  & $-36.9168$   &  $+0.5431$   & $1.949$  & $-0.064$   \\
 3530021869552726144 & $189.6879$ & $-12.1395$  & $28.96$ & $18.01$  & $1.087$  &  $-0.753$   & $0.191$  & $0.126$  & $-31.7583$   &  $-0.1585$   & $2.122$  & $-0.040$   \\
 3624655156023507968 & $197.2064$ & $ -9.4421$  & $25.77$ & $17.83$  & $0.894$  &  $-0.873$   & $0.151$  & $0.108$  & $-23.9660$   &  $+0.8866$   & $2.161$  & $-0.007$   \\
 3627914172911966848 & $197.4735$ & $ -8.1547$  & $30.60$ & $18.10$  & $0.646$  &  $-0.637$   & $0.171$  & $0.122$  & $-23.4155$   &  $+2.0801$   & $1.744$  & $ 0.128$   \\
 3619333313586245504 & $209.2637$ & $ -7.4897$  & $24.67$ & $17.69$  & $1.133$  &  $-0.843$   & $0.135$  & $0.096$  & $-11.9247$   &  $-0.1294$   & $2.583$  & $-0.029$   \\
 3619722162745300992 & $209.0226$ & $ -6.5525$  & $24.13$ & $17.59$  & $0.930$  &  $-0.867$   & $0.131$  & $0.090$  & $-11.9135$   &  $+0.8377$   & $2.419$  & $ 0.034$   \\
 3641032480502389120 & $213.9961$ & $ -5.6457$  & $24.63$ & $17.63$  & $0.917$  &  $-0.796$   & $0.111$  & $0.101$  & $ -6.9038$   &  $+0.4099$   & $2.420$  & $ 0.072$   \\
 6335641257742513920 & $227.4895$ & $ -4.7468$  & $22.82$ & $17.69$  & $1.373$  &  $-0.815$   & $0.129$  & $0.121$  & $  6.2584$   &  $-2.4580$   & $3.020$  & $-0.005$   \\
 4422749311862591104 & $236.6305$ & $ +1.0298$  & $24.25$ & $17.75$  & $1.459$  &  $-0.881$   & $0.111$  & $0.105$  & $ 16.6650$   &  $+0.4624$   & $2.934$  & $-0.167$   \\
 4409028300102799616 & $239.1512$ & $ -2.1963$  & $18.19$ & $17.55$  & $1.876$  &  $-0.871$   & $0.124$  & $0.086$  & $ 18.1559$   &  $-3.3506$   & $3.920$  & $ 0.121$   \\
 4409997347803313664 & $239.4776$ & $ +0.3137$  & $20.00$ & $17.50$  & $1.687$  &  $-0.675$   & $0.117$  & $0.118$  & $ 19.1877$   &  $-1.0388$   & $3.572$  & $ 0.212$   \\
 4412725751547859712 & $240.1723$ & $ +2.3099$  & $20.27$ & $17.38$  & $1.949$  &  $-0.765$   & $0.096$  & $0.071$  & $ 20.4236$   &  $+0.6756$   & $3.757$  & $ 0.018$   \\
 4432859557502267904 & $247.3155$ & $ +2.4327$  & $21.63$ & $17.50$  & $1.779$  &  $-0.806$   & $0.126$  & $0.094$  & $ 27.3037$   &  $-1.2277$   & $3.382$  & $-0.024$   \\
 \enddata
\end{deluxetable*}
\end{rotatetable}


\begin{deluxetable*}{rrrrrrrrr}
\tablecaption{The member RGB stars of the  new stream are listed.
    Columns from left to the right are the source ID from Gaia DR3, the coordinates ($\alpha$, $\delta$),  
    the distance $d$, the $G-$band magnitude, the proper motions and 
    their uncertainties along $\alpha$ and $\delta$, 
    the new coordinate ($\phi_1$, $\phi_2$), and the proper 
    motions along $\phi_1$ and  $\phi_2$ with Solar's movement
    corrected.}\label{tab:RGB}
    \tablehead{\\
  \colhead{source ID} &\colhead{$\alpha$} & \colhead{$\delta$} & \colhead{$\mu_\alpha^*$} & \colhead{$\mu_\delta$}  & \colhead{$RV$} & \colhead{$\sigma_{RV}$} & \colhead{$SNR_i$} & \colhead{$d_R$}   \\   
            &\colhead{$^\circ$}   & \colhead{$^\circ$}    & \colhead{mas yr$^{-1}$}& \colhead{mas yr$^{-1}$}  & \colhead{km s$^{-1}$}   & \colhead{km s$^{-1}$} &     & \colhead{kpc}}     
\startdata
  3649315067235241600 & 218.7366 & -2.0024 & 1.0637 & -0.7411 & -108.8 & 28.3 & 20.7 & 23.2\\
  4416467733212650624 & 234.7619 & -0.3649 & 1.2960 & -0.8540 &  -95.9 & 10.1 & 27.2 & 21.4\\
  4416442577587619328 & 234.5613 & -0.4576 & 1.3861 & -0.9474 &  -78.9 & -    & 28.1 & 21.4\\
  6340521616196050560 & 225.1873 & -2.0086 & 1.3030 & -0.7603 &  -84.2 &  6.8 & 23.2 & 22.3\\
  4416756492453565952 & 234.2033 &  0.3582 & 1.7631 & -0.7256 &  -76.4 & 13.0 & 22.4 & 21.4\\
  4422678633880782464 & 237.0727 &  0.7388 & 1.6171 & -0.9054 &  -92.2 &  8.4 & 21.8 & 21.2\\
  4438965596543799936 & 248.5661 &  6.8510 & 1.8232 & -0.9404 & -133.1 &  2.6 & 65.6 & 20.6 \\
 \enddata
\end{deluxetable*}

\bibliography{Gaia_SSFind_COC}{}
\bibliographystyle{aasjournal}


\end{CJK*}
\end{document}